\documentclass[10pt,english,notitlepage,a4paper,english,superscriptaddress,floatfix,longbibliography]{revtex4-1}
\usepackage[T1]{fontenc}
\usepackage{color}
\usepackage{babel}
\usepackage{amsmath}
\usepackage{amssymb}
\usepackage{graphicx}
\usepackage[unicode=true,pdfusetitle,
 bookmarks=true,bookmarksnumbered=false,bookmarksopen=false,
 breaklinks=false,pdfborder={0 0 1},backref=false,colorlinks=true]
 {hyperref}

\makeatletter

\usepackage[compat=1.1.0]{tikz-feynman}

\makeatother

\begin{document}
\title{A Density Consistency approach to the inverse Ising problem}
\author{Alfredo Braunstein}
\affiliation{Politecnico di Torino, Corso Duca Degli Abruzzi 24, Torino, Italy }
\affiliation{Italian Institute for Genomic Medicine, Via Nizza 52, Torino, Italy}
\affiliation{Collegio Carlo Alberto, Piazza Arbarello 8, 10122 Torino, Italy }
\affiliation{INFN Sezione di Torino, Via P. Giuria 1, I-10125 Torino, Italy}
\author{Giovanni Catania}
\email{giovanni.catania@polito.it}
\affiliation{Politecnico di Torino, Corso Duca Degli Abruzzi 24, Torino, Italy }
\author{Luca Dall'Asta}
\affiliation{Politecnico di Torino, Corso Duca Degli Abruzzi 24, Torino, Italy }
\affiliation{Collegio Carlo Alberto, Piazza Arbarello 8, 10122 Torino, Italy }
\affiliation{INFN Sezione di Torino, Via P. Giuria 1, I-10125 Torino, Italy}
\author{Anna Paola Muntoni}
\affiliation{Politecnico di Torino, Corso Duca Degli Abruzzi 24, Torino, Italy }
\begin{abstract}
We propose a novel approach to the inverse Ising problem  which employs the recently introduced Density Consistency approximation (DC) to determine the model parameters (couplings and external fields) maximizing the likelihood of given empirical data. This method allows for closed-form expressions of the inferred parameters as a function of the first and second empirical moments. Such expressions have a similar structure to the small-correlation expansion derived in Ref.~\cite{sessak_small-correlation_2009}, of which they provide an improvement in the case of non-zero magnetization at low temperatures, as well as in presence of random external fields. The present work provides an extensive comparison with most common inference methods used to reconstruct the model parameters in several regimes, i.e. by varying both the network topology and the distribution of fields and couplings. The comparison shows that no method is uniformly better than every other one, but DC appears nevertheless as one of the most accurate and reliable approaches to infer couplings and fields from first and second moments in a significant range of parameters.
\end{abstract}
\maketitle

\section{Introduction and related work}

The main goal of statistical physics is to predict the macroscopic behavior of a system knowing the microscopic interactions acting on its elements, formally described by the Hamiltonian. One of the most studied problems in statistical mechanics, the \textit{direct} Ising problem, consists of determining some equilibrium observables, such as the magnetization, of a system of magnets (spins) characterized by pairwise interactions. In the present work, we will refer to the Ising model as any binary spin model with arbitrary sets of pairwise couplings and external fields, defined on a certain topology. A huge amount of problems of practical interest can be formally re-phrased in terms of an Ising problem and, furthermore, the recent advances of empirical capabilities and enhanced data availability have drawn the attention to the \textit{inverse} Ising problem (IIP) in which, from given empirical measurements, we attempt to learn the microscopic properties of the target systems, that is the pairwise couplings and external fields of the Hamiltonian. 
Applications can be found in several research domains: in neuroscience, to reconstruct neural connections from time series of neural spikes \citep{CoccoLeibler,Tyrcha_2013}, in molecular biology, to reconstruct gene regulatory networks \citep{Lezon19033,Locasale_dynamicsignaling}, or even in econophysics, to analyze stock market data and predict the behavior of financial markets \citep{borysov_us_2015,BURY20131375}.
%\textcolor{green}{AB: maybe talk about heterogeinity here, as Ising by itself could be confusing without it} 
%The Inverse Ising Problem (I.I.P.) attempts to infer the model parameters, represented by pairwise couplings and external fields, given some empirical measurements about equilibrium configurations.

%In the last two decades, the I.I.P. has found many applications in several research domains, also thanks to recent advances of empirical capabilities and enhanced data availability: for instance, applications can be found in neuroscience, to reconstruct neural connections from times series of neuro spikes \citep{CoccoLeibler,Tyrcha_2013}; in molecular biology, to reconstruct gene regulatory networks \citep{Lezon19033,Locasale_dynamicsignaling}. Finally, applications have been designed in econophysics, to analyze stock market data and predict the behaviour of financial markets \citep{borysov_us_2015,BURY20131375}.

%Within a Bayesian framework, the IIP can be formally re-phrased as the problem of determining the set of parameters, i.e. couplings and external fields, given the data (often encoded in all the possible empirical first and second moments of the data distribution) which maximize the posterior probability, eventually including some prior knowledge about the topology (i.e. the sparsity of the graph). 

The IIP can be formally re-phrased as the Bayesian problem of determining the set of unknown parameters, i.e. the couplings and the external fields, which are able to reproduce the given observables, often encoded in the empirical first and second moments of the data distribution. Its solution consists of those parameters which maximize the posterior probability, the so-called MAP (maximum a posteriori) estimator. In cases where no prior knowledge about the model parameters is assumed, i.e. graph topology or even a regularization term, this estimator coincides with the set of parameters that maximizes the Likelihood function \citep{mackaybook}. However, an exact solution is possible only when the number of empirical configurations goes to infinity, and, even in that case, the computational cost to find the maximum likelihood estimator typically scales exponentially with the system size.

For this reason, in the statistical physics community a lot of effort
has been devoted to the analysis of different approximations to solve
the IIP. For instance, Mean-Field (MF) \citep{peterson,kappen_ortiz}
and TAP \citep{TAP_first,tanaka_tap} approximations lead to simple, closed-form expressions for the couplings and the fields, obtained by using Linear Response theory.
Another way of approaching the problem is by using the Bethe Approximation
\citep{mezard2009information,yedidia2003understanding}: in this setting,
a closed solution of the maximum likelihood equations can be found
and it is known in the literature as the Independent Pair (IP) approximation
\citep{roudi_statistical_2009,roudi_ising_2009}. However, even in
the limit of infinite samples the resulting expression turns out to
be exact only if the topology is known and the interactions graph is acyclic. The methods described so far can be formulated in terms of variational approximations of the true free energy: in particular, MF and TAP represent respectively the first-order and second-order approximation of the well-known Plefka Expansion \citep{plefka_convergence_1982}, obtained by expanding
the Gibbs Free Energy at fixed magnetization with respect to a perturbation parameter associated with the interaction terms in the Hamiltonian. In this context, the IP approximation is obtained by summing all the two-spin diagrams in the Plefka expansion, which
leads to the Bethe Free Energy at fixed magnetizations and correlations.

Further improvements to the Bethe Approximation have been developed
to estimate couplings by means of the Linear Response theory: an example is given
by the Susceptibility Propagation algorithm, developed first in \citep{welling_approximate_2003} and afterwards in \citep{SPMezardMora} specifically for the Inverse Ising Problem, and re-derived analytically in \citep{SP_Ricci_Tersenghi,Nguyen_2012}.
Other methods have been proposed to take into account the effect
of loops in the estimation of couplings and fields, as it has been
done in \citep{sessak_small-correlation_2009}. 

All the methods previously described give an analytic expression of the model
parameters only in terms of first and second moments; further developments  where higher-order correlations are considered have been shown in \citep{ACE_cocco_monasson1,ACE_cocco_monasson2}, with the Adaptative Cluster Expansion. Another way of (implicitly) including high-order correlations is by using Pseudo-Likelihood (PL) Maximization, first introduced in \citep{Besag_PL} and then re-discovered in the statistical physics community \citep{Aurell_Ekeberg_PL}. PL is widely considered as one of the best inference methods to solve the IIP \citep{nguyen_inverse_2017}, providing very accurate estimates of model parameters in a running time that scales quadratically and linearly with respect to the number of spins and the number of empirical samples, respectively. For completeness, we will include a comparison also with PL, but it should be remembered that this method includes extra information (namely the full set of samples) not available to the mean-field based methods just presented.
%Therefore, we will compare our results also to PL but keeping in mind that the amount of information used by PL is extremely larger than all the other Mean-Field like methods we are going to describe.
%\textcolor{green}{AB: For completeness, we will include a comparison also with PL, but it should be remembered that this method includes extra information (namely the full set of samples) not available to the rest of them. -- forse includere un disclaimer simile nelle conclusioni?}

We refer to \citep{nguyen_inverse_2017} for an exhaustive analysis
of all these methods. We mention that some of them can be easily extended
to the inference of models of multi-state variables, like the Potts variables, that has a lot of interesting applications in computational biology, in particular in the prediction of the three-dimensional protein structures from co-evolutionary related sequences: see for instance \citep{Cocco_Proteins} and references therein.

The purpose of this work is to describe another way of solving the
IIP using Density Consistency (DC), a novel class of approximation schemes
introduced in \citep{braunstein_loop_2019} to compute approximately
marginal probability distributions over discrete Markov Random Fields. Its derivation is similar to the well-known Expectation Propagation algorithm \citep{minka_expectation_2001,opper_adaptive_2001} with the main difference of relying on a modified consistency condition allowing for the exact computation of the marginal densities in the case of acyclic graphs. In this sense, it represents a generalization of the Bethe Approximation \citep{mezard2009information,yedidia2003understanding} as the DC estimate of equilibrium observables, in case of loopy graphs, are significantly improved. 

The paper is structured as follows: in Section \ref{sec:Problem-Setup}
we define the inverse Ising problem within a standard Bayesian approach.
Section \ref{sec:DC_EP_section} describes the Density Consistency
method in the context of pairwise models while Section \ref{sec:ML_solution_DC_EP} presents the main analytic results of this work, i.e. the expression of inferred couplings and fields under DC approximation. Section \ref{sec:Results} shows the results of numerical simulations on synthetic data, and the comparison to other inference methods. Section \ref{sec:Conclusions} summarizes the main findings and presents the future directions for forthcoming works.

\section{Problem Setup\label{sec:Problem-Setup}}

We consider an Ising model defined on an arbitrary graph $G=\left(V,E\right)$,
where each node represents a binary variable (spin) $\sigma_{i}=\left\{ -1,+1\right\} ,i\in V=\left\{ 1,\ldots,N\right\} $.
The model is parametrized by a set of symmetric couplings $J_{ij},\left(i,j\right)\in E$
encoding pairwise interactions between neighbour spins and by a set
of external local fields $\left\{ h_{i}\right\} _{i=1}^{N}$, each
one acting on a different variable. At a certain inverse temperature
$\beta$, the equilibrium statistics of the model is described by
the Boltzmann measure
\begin{equation}
p\left(\boldsymbol{\sigma}|\boldsymbol{h},\boldsymbol{J}\right)=\frac{1}{Z\left(\boldsymbol{h},\boldsymbol{J}\right)}\exp\left[\sum_{i<j}\beta J_{ij}\sigma_{i}\sigma_{j}+\sum_{i}\beta h_{i}\sigma_{i}\right]=\frac{1}{Z\left(\boldsymbol{h},\boldsymbol{J}\right)}\exp\left[-\beta H\left(\boldsymbol{\sigma}|\boldsymbol{h},\boldsymbol{J}\right)\right]\label{eq:ising_pdf}
\end{equation}
where $H\left(\boldsymbol{\sigma}|\boldsymbol{h},\boldsymbol{J}\right)=-\sum_{i<j}J_{ij}\sigma_{i}\sigma_{j}-\sum_{i}h_{i}\sigma_{i}$
is the Ising Hamiltonian. Here the summation runs over all distinct
pairs $i<j$, and couplings are assumed to be zero for a non-existing
edge, i.e. $J_{ij}=0\;\forall i,j:\left(i,j\right)\notin E$. 

For sake of simplicity and motivated by the impossibility in the inverse
problem of determining the temperature, we will absorb $\beta$ into the model parameters, namely $\beta h_{i}\to h_{i}$ and $\beta J_{ij}\to J_{ij}$. The prefactor in (\ref{eq:ising_pdf}) denotes the inverse of the partition function
\begin{equation}
Z\left(\boldsymbol{h},\boldsymbol{J}\right)=\sum_{\boldsymbol{\sigma}}\exp\left[\sum_{i<j} J_{ij}\sigma_{i}\sigma_{j}+\sum_{i} h_{i}\sigma_{i}\right],\label{eq:partition_function_Z}
\end{equation}
where $\boldsymbol{\sigma}=\left\{ \sigma_{i}\right\} _{i=1}^{N}$.

From an information-theory perspective, (\ref{eq:ising_pdf}) can
also be computed as the maximum entropy distribution at fixed first
and second empirical moments \citep{nguyen_inverse_2017}. Typically,
one is interested in computing expectation values of (\ref{eq:ising_pdf}),
at given parameters: this is known as the \emph{direct} problem.
On the other hand, in the \emph{inverse} problem we are interested
in estimating the model parameters by knowing the statistics of the underlying
distribution.

%\subsubsection{Maximum Likelihood}
Suppose we are provided with a certain number $M$ of samples drawn
from (\ref{eq:ising_pdf}). We denote with $\left\{ \boldsymbol{\sigma}^{\mu}\right\} _{\mu=1}^{M}$
the set of configurations, where each one is a particular realization
of (\ref{eq:ising_pdf}). By means of Bayes Theorem \citep{mackaybook},
we can express the \emph{posterior} probability of the model parameters
$\boldsymbol{\theta}=\left(\boldsymbol{h},\boldsymbol{J}\right)$,
given the data:
\begin{equation}
p\left(\boldsymbol{\theta}|\left\{ \boldsymbol{\sigma}^{\mu}\right\} _{\mu=1}^{M}\right)=\frac{p\left(\left\{ \boldsymbol{\sigma}^{\mu}\right\} _{\mu=1}^{M}|\boldsymbol{\theta}\right)p\left(\boldsymbol{\theta}\right)}{p\left(\left\{ \boldsymbol{\sigma}^{\mu}\right\} _{\mu=1}^{M}\right)}\label{eq:Bayestheorem}
\end{equation}
where $p\left(\boldsymbol{\theta}\right)$ is the prior distribution
and $p\left(\left\{ \boldsymbol{\sigma}^{\mu}\right\} _{\mu=1}^{M}|\boldsymbol{\theta}\right)$
is the likelihood function. The latter represents the probability
that the set of configurations $\left\{ \boldsymbol{\sigma}^{\mu}\right\} _{\mu=1}^{M}$
has been drawn from the starting distribution (\ref{eq:ising_pdf})
with parameters $\boldsymbol{\theta}$, and it is interpreted as a
function of $\boldsymbol{\theta}$. The prior distribution can take
into account some additional information about the model parameters:
for instance, knowledge about the graph sparsity can be encoded by
means of a $L^{p}$-norm prior, also known as regularization (we do not address this point within this work). 

Typically, one attempts to find the best set of parameters $\boldsymbol{\theta}^{*}$ maximizing the posterior distribution, and the resulting estimator is known as maximum-a-posteriori (MAP). Very often, no prior knowledge about model parameters is assumed (i.e. $p\left(\boldsymbol{\theta}\right)$ is uniformly distributed): as a consequence, the MAP estimator coincides with the maximum likelihood (ML) point. For the Ising model (\ref{eq:ising_pdf}), under the assumption that samples are independent and identically distributed (i.i.d), the likelihood function can be factorized over the samples:
\begin{align}
\mathcal{L}\left(\boldsymbol{h},\boldsymbol{J}\right)\hat{=}\frac{1}{M}\log p\left(\left\{ \boldsymbol{\sigma}^{\mu}\right\} _{\mu=1}^{M}|\boldsymbol{\theta}\right) & = \frac{1}{M} \sum_{\mu = 1}^{M} \log p\left( \boldsymbol{\sigma}^{\mu} |\boldsymbol{\theta}\right) \nonumber \\& = \sum_{i}h_{i}\frac{1}{M}\sum_{\mu = 1}^{M}\sigma_{i}^{\mu}+\sum_{i<j}J_{ij}\frac{1}{M}\sum_{\mu=1}^{M}\sigma_{i}^{\mu}\sigma_{j}^{\mu}-\log Z\left(\boldsymbol{h},\boldsymbol{J}\right)\nonumber \\
 & =\sum_{i}h_{i}m_{i}+\sum_{i<j}J_{ij}\chi_{ij}-\log Z\left(\boldsymbol{h},\boldsymbol{J}\right)\label{eq:loglikelihood}
\end{align}
where $Z$ is the partition function defined in (\ref{eq:partition_function_Z})
and $\mathcal{L}$ the log-likelihood function (normalized
over $M$). In (\ref{eq:loglikelihood}), $m_{i}$ and $\chi_{ij}$
denote the first and second empirical moments, respectively:
\begin{align}
m_{i} & \hat{=}\frac{1}{M}\sum_{\mu=1}^{M}\sigma_{i}^{\mu},\qquad\chi_{ij}\hat{=}\frac{1}{M}\sum_{\mu=1}^{M}\sigma_{i}^{\mu}\sigma_{j}^{\mu}\label{eq:exp_moments}
\end{align}

The maximum likelihood estimator is found by setting to $0$ the derivatives
of $\mathcal{L}$ with respect to the model parameters $\boldsymbol{\theta}$,
which in turn implies that the first and second empirical moments have
to be equal to the equilibrium expectation values over (\ref{eq:ising_pdf})
\begin{equation}
m_{i}=\langle\sigma_{i}\rangle_{p}\;\forall i,\qquad\chi_{ij}=\langle\sigma_{i}\sigma_{j}\rangle_{p}\;\forall i,j\label{eq:Boltzmann_Learning}
\end{equation}
where $\langle f\left(\boldsymbol{\sigma}\right)\rangle$ denotes
the expectation value over the Boltzmann measure \eqref{eq:ising_pdf}, $\langle f\left(\boldsymbol{\sigma}\right)\rangle=\sum_{\boldsymbol{\sigma}}f\left(\boldsymbol{\sigma}\right)p\left(\boldsymbol{\sigma}\right)$.
In the following, we will denote with $\boldsymbol{m}$ the vector of empirical magnetizations and with $\boldsymbol{C}=\boldsymbol{\chi}-\boldsymbol{m}\boldsymbol{m}^{t}$
the empirical covariance matrix. 

In principle, the maximum likelihood point can be found by means of a gradient ascent algorithm, known as Boltzmann machine learning \citep{AckleyDavidH1985Alaf}: since the log-likelihood is a convex function, the maximum can be found exactly in the limit where the number of samples goes to infinity. However, computing expectation values over the equilibrium distribution (\ref{eq:ising_pdf}) scales exponentially with the system size (i.e. $O\left(2^{N}\right)$): for very small systems the computation can be performed explicitly but for larger (and often more interesting) systems one may rely on a Monte Carlo Markov Chain (MCMC) estimate that, if performed at each update of the parameters, can be time demanding. 
We propose in the following an analytic expression of the target model parameters such that the maximum likelihood equations in (\ref{eq:Boltzmann_Learning}) is satisfied, when the model statistics is evaluated by the Density Consistency approximation \citep{braunstein_loop_2019}.

%Giova: hai ragione, io avevo scritto nella sezione successiva che la soluzione e\' analitica se usi DC, però forse e\' meglio scriverlo gia\' adesso, cosi\' da non avere il dubbio. Anna: Esatto!

%Samples can be generated forinstance according to MonteCarlo methods; alternatively, if the system size $N$ is sufficiently small, it is still possible to perform an exact trace over all the $2^{N}$ configurations. (TO BE PUT IN THE RESULTS SECTION)

\section{Density Consistency\label{sec:DC_EP_section}}

In this section we are going to give a detailed derivation of the Density Consistency method. In general, it applies to factorized distributions of binary variables, that may encode high-order interactions between group of neighbouring spins (distributions of this kind can be represented in terms of factor graphs \citep{mezard2009information,Clifford90}).
For the sake of simplicity, in the following we will recall the derivation of DC approximation applied to distributions of binary variables with pairwise interactions (i.e. Ising-like). We start from (\ref{eq:ising_pdf}) and we rewrite it as a factorized distribution of continuous variables
$\left\{ x_{i}\right\} _{i=1}^{N}$:
\begin{equation}
p\left(\boldsymbol{x}\right)=\frac{1}{Z}\prod_{i<j}\psi_{ij}\left(x_{i},x_{j}\right)\prod_{i}\Delta_{i}\left(x_{i}\right)\label{eq:pdf_continuos}
\end{equation}
where $\psi_{ij}\left(x_{i},x_{j}\right)=\exp\left[J_{ij}x_{i}x_{j}+h_{i}^{\left(ij\right)}x_{i}+h_{j}^{\left(ij\right)}x_{j}\right]$,
$\Delta_{i}=\frac{1}{2}\left[\delta\left(x_{i}-1\right)+\delta\left(x_{i}+1\right)\right]$
for $\delta\left(x\right)$ is the Dirac delta function. The set of constraints $\left\{ \Delta_{i}\right\} _{i=1}^{N}$ ensures that the distribution is discrete over $\left\{ -1,1\right\} $ for all variables $x_{i}$, so that (\ref{eq:pdf_continuos}) coincides with (\ref{eq:ising_pdf}).
In this notation, $h_{i}^{\left(ij\right)}$ (resp. $h_{j}^{\left(ij\right)}$)
denotes a certain fraction of the local field $h_{i}$ (resp. $h_{j}$)
contained into $\psi_{ij}$, with the constraint that $\sum_{j\neq i}h_{i}^{\left(ij\right)}=h_{i},\;\forall i$. \\
DC is based on a refined Gaussian approximation of (\ref{eq:pdf_continuos}), that allows for an analytic marginalization over any subset of variables.
For each edge $\left(ij\right)$, we associate an approximate
Gaussian factor $\phi_{ij}\left(x_{i},x_{j}\right)$ with the
``true'' factor $\psi_{ij}\left(x_{i},x_{j}\right)$ in (\ref{eq:pdf_continuos}), formally written as:
\begin{equation}
\phi_{ij}\left(x_{i},x_{j}\right)=\exp\left[-\frac{1}{2}\left(x_{i},x_{j}\right)\boldsymbol{\Gamma}^{\left(ij\right)}\left(x_{i},x_{j}\right)^{t}+\left(x_{i},x_{j}\right)\boldsymbol{\gamma}^{\left(ij\right)}\right].\label{eq:phi_ij}
\end{equation}
Here $\boldsymbol{\Gamma}^{\left(ij\right)}$ is a $2\times2$ matrix,
$\boldsymbol{\gamma}^{\left(ij\right)}$ is a $2-$components column
vector, both to be determined within the approximation, and parametrized as follows:
\begin{equation}
\boldsymbol{\Gamma}^{\left(ij\right)}=\begin{pmatrix}\Gamma_{ii}^{\left(ij\right)} & \Gamma_{ij}^{\left(ij\right)}\\
\Gamma_{ij}^{\left(ij\right)} & \Gamma_{jj}^{\left(ij\right)}
\end{pmatrix},\qquad\boldsymbol{\gamma}^{\left(ij\right)}=\begin{pmatrix}\gamma_{i}^{\left(ij\right)}\\
\gamma_{j}^{\left(ij\right)}
\end{pmatrix}\label{eq:Gamma_and_gamma_params}
\end{equation}
Taking the product of $\phi_{ij}$ over all distinct pairs leads to
a multivariate Gaussian distribution $q\left(\boldsymbol{x}\right)$
over the full set of variables. We express it in standard form in terms of first and second moments:
\begin{align}
q\left(\boldsymbol{x}\right) & \propto\prod_{i<j}\phi_{ij}\left(x_{i},x_{j}\right)\propto\exp\left[-\frac{1}{2}\left(\boldsymbol{x}-\boldsymbol{\mu}\right)^{t}\boldsymbol{\Sigma}^{-1}\left(\boldsymbol{x}-\boldsymbol{\mu}\right)\right],\label{eq:dc_fullgauss}
\end{align}
which is defined up to a normalization constant. The Gaussian moments
$\boldsymbol{\mu}\in\mathbb{R}^{N}$ and $\boldsymbol{\Sigma}\in\mathbb{R}^{N\times N}$
are related to the set of parameters $\left\{ \left(\boldsymbol{\gamma}^{\left(ij\right)},\boldsymbol{\Gamma}^{\left(ij\right)}\right)\right\}_{i<j} $ by
the following relations:
\begin{equation}
\left(\boldsymbol{\Sigma}^{-1}\right)_{ij}=\begin{cases}
\Gamma_{ij}^{\left(ij\right)} & \text{if }i\ne j\\
\sum_{k\ne i}\Gamma_{ii}^{\left(ik\right)} & \text{if }i=j
\end{cases},\qquad\left(\boldsymbol{\Sigma}^{-1}\boldsymbol{\mu}\right)_{i}=\sum_{k\ne i}\gamma_{i}^{\left(ik\right)}\label{eq:gauss_moments_wrt_params}
\end{equation}
For each edge $\left(ij\right)$, we define a ``tilted'' distribution
$q^{\left(ij\right)}\left(\boldsymbol{x}\right)$ obtained from $q\left(\boldsymbol{x}\right)$
by replacing the Gaussian factor $\phi_{ij}$ with the true factor
$\psi_{ij}$
\begin{equation}
q^{\left(ij\right)}\left(\boldsymbol{x}\right)\propto\prod_{\overset{k<l}{(kl)\neq(ij)}}\phi_{kl}\Psi_{ij}\propto q\left(\boldsymbol{x}\right)\frac{\Psi_{ij}}{\phi_{ij}},\label{eq:qtilted_ij}
\end{equation}
%\[
%q^{\left(ij\right)}\left(\boldsymbol{x}\right)\propto g^{\backslash\left(ij\right)}\left(\boldsymbol{x}\right)\Psi_{ij}\qquad q\left(\boldsymbol{x}\right)\propto g^{\backslash\left(ij\right)}\left(\boldsymbol{x}\right)\phi_{ij}
%\]
where $\Psi_{ij}=\psi_{ij}\Delta_{i}\Delta_{j}$. In this way we encode
the information about the discrete nature of variables $i$ and $j$
into the tilted distribution. The main idea is that, for each edge,
its corresponding tilted distribution will be a good estimator of
the marginals (or equivalently, first and second moments) of the true
distribution, since the factor $\psi_{ij}$ is correctly included
in (\ref{eq:qtilted_ij}). Computation of marginals over (\ref{eq:qtilted_ij})
is possible because all other variables are encoded into a Gaussian
factor, that allows for an analytical marginalization. The marginal
distribution of (\ref{eq:qtilted_ij}) over $\left(ij\right)$ can
therefore be written as
\begin{equation}
q^{\left(ij\right)}\left(x_{i},x_{j}\right)=\int d\boldsymbol{x}_{\backslash i,j}q^{\left(ij\right)}\left(\boldsymbol{x}\right)\propto g^{\backslash\left(ij\right)}\left(x_{i},x_{j}\right)\Psi_{ij}\left(x_{i},x_{j}\right),\label{eq:marginal_tilted_qij}
\end{equation}
where $g^{\backslash ij}$ is a Gaussian cavity distribution,
marginalized over all the variables except $i,j$: it carries out
an effective interaction between spins $i$ and $j$ coming from all
the cycles in the graph. We express it as:
\[
g^{\backslash\left(ij\right)}\left(x_{i},x_{j}\right)\propto\exp\left[-\frac{1}{2}\left(x_{i},x_{j}\right)\boldsymbol{S}^{\left(ij\right)}\left(x_{i},x_{j}\right)^{t}+\left(x_{i},x_{j}\right)\boldsymbol{y}^{\left(ij\right)}\right],
\]
where, again, $\boldsymbol{S}^{\left(ij\right)}$ is a $2\times2$
matrix, $\boldsymbol{y}^{\left(ij\right)}$ is a $2-$components column
vector. It is straightforward to see that these terms are functions of the full Gaussian moments $\boldsymbol{\mu},\boldsymbol{\Sigma}$, as shown below:
\begin{align}
y_{i}^{\left(ij\right)} & =\frac{\Sigma_{jj}\mu_{i}-\Sigma_{ij}\mu_{j}}{\Sigma_{ii}\Sigma_{jj}-\Sigma_{ij}^{2}}-\gamma_{i}^{\left(ij\right)}\qquad y_{j}^{\left(ij\right)}=\frac{-\Sigma_{ij}\mu_{i}+\Sigma_{ii}\mu_{j}}{\Sigma_{ii}\Sigma_{jj}-\Sigma_{ij}^{2}}-\gamma_{j}^{\left(ij\right)}\label{eq:cav_field_ij}
\end{align}
\begin{equation}
S_{ij}^{^{\left(ij\right)}}=\frac{-\Sigma_{ij}}{\Sigma_{ii}\Sigma_{jj}-\Sigma_{ij}^{2}}-\left(\boldsymbol{\Sigma}^{-1}\right)_{ij}\label{eq:cav_coupling_Sij}.
\end{equation}
The first and second moments of each marginal tilted distribution in (\ref{eq:marginal_tilted_qij}), can now be analytically computed by summing over $(x_{i},x_{j})\in \{+1,-1\}^2$:
\begin{subequations}
\label{tilted_moments_all}
\begin{align}
\langle x_{i}\rangle_{q^{\left(ij\right)}} & =\tanh\left[a_{i}^{\left(ij\right)}+\text{atanh}\left(\tanh b^{\left(ij\right)}\tanh a_{j}^{\left(ij\right)}\right)\right],\label{eq:mi_qij}\\
\langle x_{j}\rangle_{q^{\left(ij\right)}} & =\tanh\left[a_{j}^{\left(ij\right)}+\text{atanh}\left(\tanh b^{\left(ij\right)}\tanh a_{i}^{\left(ij\right)}\right)\right],\label{eq:mj_qij}\\
\langle x_{i}x_{j}\rangle_{q^{\left(ij\right)}} & =\tanh\left[b^{\left(ij\right)}+\text{atanh}\left(\tanh a_{i}^{\left(ij\right)}\tanh a_{j}^{\left(ij\right)}\right)\right],\label{eq:mij_qij}
\end{align}
\end{subequations}
where $a_{i}^{\left(ij\right)}=h_{i}^{\left(ij\right)}+y_{i}^{\left(ij\right)}$,
$a_{j}^{\left(ij\right)}=h_{j}^{\left(ij\right)}+y_{j}^{\left(ij\right)}$,
$b^{\left(ij\right)}=J_{ij}-S_{ij}^{\left(ij\right)}$. Notice that
the effect of the cavity distribution is to add an effective coupling
$-S_{ij}^{\left(ij\right)}$(i.e. the off-diagonal term of the matrix
$\boldsymbol{S}^{\left(ij\right)}$), together with the cavity fields $y_{i}^{\left(ij\right)},y_{j}^{\left(ij\right)}$
.

\subsubsection{Closures and update scheme}

If the model is known and we are interested in computing the marginal probabilities of (\ref{eq:ising_pdf}), i.e. we are interested in solving the direct problem, the Gaussian parameters $\left\{ \left(\boldsymbol{\gamma}^{\left(ij\right)},\boldsymbol{\Gamma}^{\left(ij\right)}\right)\right\}_{i<j}$ can be iteratively determined requiring that a certain \textit{consistency} condition is satisfied  between the full Gaussian distribution $q\left(\boldsymbol{x}\right)$ and the set of tilted distributions $\left\{ q^{\left(ij\right)}\left(\boldsymbol{x}\right)\right\}_{i<j} $ associated with each edge of the corresponding factor graph.

Among all the possible choices, in \citep{braunstein_loop_2019} the authors found that any scheme satisfying
$\frac{\mu_{i}}{\Sigma_{ii}}=\text{atanh}\langle x_{i}\rangle_{q^{\left(ij\right)}}$
gives exact marginalization on acyclic graphs (trees). They then constructed a set of reasonable matching conditions to close the set of update equations (note that in each approximate factor $\phi_{ij}\left(x_{i},x_{j}\right)$ five parameters have to be determined):
\begin{subequations}
\label{DCclosure_all}
\begin{align}
\mu_{i} & =\langle x_{i}\rangle_{q^{\left(ij\right)}}\label{eq:DCclosure_mi}\\
\Sigma_{ii} & =\frac{\langle x_{i}\rangle_{q^{\left(ij\right)}}}{\text{atanh}\langle x_{i}\rangle_{q^{\left(ij\right)}}}\label{eq:DCclosure_sii}\\
\Sigma_{ij} & =\left(\langle x_{i}x_{j}\rangle_{q^{\left(ij\right)}}-\langle x_{i}\rangle_{q^{\left(ij\right)}}\langle x_{j}\rangle_{q^{\left(ij\right)}}\right)\sqrt{\frac{\Sigma_{ii}\Sigma_{jj}}{\left(1-\langle x_{i}\rangle_{q^{\left(ij\right)}}^{2}\right)\left(1-\langle x_{j}\rangle_{q^{\left(ij\right)}}^{2}\right)}}\label{eq:DCclosure_sij}
\end{align}
\end{subequations}
for all distinct pairs $i<j$. 
%In principle, there are infinite possible choices to impose consistency conditions between the Gaussian distribution and the set of tilted distributions: in particular, we found that any scheme satisfying $\frac{\mu_{i}}{\Sigma_{ii}}=\text{atanh}\langle x_{i}\rangle_{q^{\left(ij\right)}}$ gives exact marginalization on trees. 
The parameter $\eta\in\left[0,1\right]$
in (\ref{eq:DCclosure_sij}) acts an interpolation between a full
DC solution $\eta=1$ and a BP solution $\eta=0$: indeed, setting
$\eta=0$ is equivalent to assume that the cavity distribution is
factorized over the edges, and it gives the BP fixed point on any topology \citep{braunstein_loop_2019}.A detailed derivation of the above equations and its connection to Belief Propagation fixed points is discussed in Appendix \ref{app:DCeqs}.

For reasons that will be clear in the following section, we present another update scheme, directly
inspired by standard EP implementations, that is obtained by imposing first and second moments matching between $q\left(\boldsymbol{x}\right)$ and each $q^{\left(ij\right)}\left(\boldsymbol{x}\right)$:
\begin{subequations}
\label{EPclosure_all}
\begin{align}
\mu_{i} & =\langle x_{i}\rangle_{q^{\left(ij\right)}}\label{eq:EPclosure_mi}\\
\Sigma_{ii} & =1-\langle x_{i}\rangle_{q^{\left(ij\right)}}^{2}\label{eq:EPclosure_sii}\\
\Sigma_{ij} & =\langle x_{i}x_{j}\rangle_{q^{\left(ij\right)}}-\langle x_{i}\rangle_{q^{\left(ij\right)}}\langle x_{j}\rangle_{q^{\left(ij\right)}}\label{eq:EPclosure_sij}
\end{align}
\end{subequations}
again, for all distinct pairs $i<j$. In the following, we will refer
to \eqref{DCclosure_all} as ``DC'' closure and \eqref{EPclosure_all}
as ``EP'' closure.

\section{Analytic solution of the Maximum Likelihood Equations\label{sec:ML_solution_DC_EP}}

We have mentioned in the previous section that we consider the set of tilted distributions as those that better approximate the true marginal probabilities. The ML condition  (\ref{eq:Boltzmann_Learning}) for the \textit{inverse} problem implies the matching between the model and the empirical first and second order statistics. Therefore, in order to solve the inverse problem we will require that the statistics computed according to our tilted densities must match the empirical statistics as
\begin{subequations}
\label{Boltzmann_learning_all}
\begin{align}
\langle x_{i}\rangle_{q^{\left(ij\right)}} & =m_{i}\label{eq:BLearningDC1}\\
\langle x_{j}\rangle_{q^{\left(ij\right)}} & =m_{j}\label{eq:BLearningDC2}\\
\langle x_{i}x_{j}\rangle_{q^{\left(ij\right)}} & =C_{ij}+m_{i}m_{j}\label{eq:BLearningDC3}
\end{align}
\end{subequations}
Note that the l.h.s of (\ref{eq:BLearningDC3}) depends on the unknown parameters $h_{i}^{\left(ij\right)}$,$h_{j}^{\left(ij\right)}$,$J_{ij}$,
for $\forall i<j$: since the r.h.s is determined by the data, we can solve (\ref{eq:BLearningDC3}) w.r.t.  to the couplings and the external fields allowing us to solve the inverse Ising problem within the DC approximation scheme.
%Finally, the overall external field acting on $x_{i}$ will be computed as $h_{i}=\sum_{j\neq i}h_{i}^{\left(ij\right)}$.
First notice that, by putting together  \eqref{Boltzmann_learning_all}
and \eqref{tilted_moments_all}, the system of three equations can be solved with respect to the parameters $a_{i}^{(ij)}$, $a_{j}^{(ij)}$, $b^{(ij)}$ at fixed moments $m_{i}$, $m_{j}$, $C_{ij}$:
\begin{subequations}

\label{IP_first}
\begin{align}
a_{i}^{\left(ij\right)}	&= h_{i}^{(ij)} + y_{i}^{(ij)}= \frac{1}{4}\log\left\{ \frac{\left[\left(1+m_{i}\right)\left(1+m_{j}\right)+C_{ij}\right]\left[\left(1+m_{i}\right)\left(1-m_{j}\right)-C_{ij}\right]}{\left[\left(1-m_{i}\right)\left(1+m_{j}\right)-C_{ij}\right]\left[\left(1-m_{i}\right)\left(1-m_{j}\right)+C_{ij}\right]}\right\}\label{IP_ai_first}\\
a_{j}^{\left(ij\right)}&	=h_{j}^{(ij)} + y_{j}^{(ij)}=\frac{1}{4}\log\left\{ \frac{\left[\left(1+m_{i}\right)\left(1+m_{j}\right)+C_{ij}\right]\left[\left(1-m_{i}\right)\left(1+m_{j}\right)-C_{ij}\right]}{\left[\left(1+m_{i}\right)\left(1-m_{j}\right)-C_{ij}\right]\left[\left(1-m_{i}\right)\left(1-m_{j}\right)+C_{ij}\right]}\right\}\label{IP_aj_first}\\
b^{\left(ij\right)}	&=  J_{ij}-S_{ij}^{(ij)}= \frac{1}{4}\log\left\{ \frac{\left[\left(1+m_{i}\right)\left(1+m_{j}\right)+C_{ij}\right]\left[\left(1-m_{i}\right)\left(1-m_{j}\right)+C_{ij}\right]}{\left[\left(1+m_{i}\right)\left(1-m_{j}\right)-C_{ij}\right]\left[\left(1-m_{i}\right)\left(1+m_{j}\right)-C_{ij}\right]}\right\}\label{IP_b_first}.
\end{align}
\end{subequations}
If one neglects the cavity parameters, a direct expression of the inferred parameters follows from \eqref{IP_first} and it is known as the Independent Pair approximation (IP) \citep{roudi_statistical_2009,roudi_ising_2009}: here, the couplings and the external fields are inferred by assuming that each pair of spin is independent on the others. In this case, the expression of the couplings follows directly from \eqref{IP_b_first}, and the fields $h_{i}$ can be reconstructed by summing over $j\neq i$:

\begin{align}
J_{ij}^{IP} & =\frac{1}{4}\log\frac{\left[\left(1+m_{i}\right)\left(1+m_{j}\right)+C_{ij}\right]\left[\left(1-m_{i}\right)\left(1-m_{j}\right)+C_{ij}\right]}{\left[\left(1+m_{i}\right)\left(1-m_{j}\right)-C_{ij}\right]\left[\left(1-m_{i}\right)\left(1+m_{j}\right)-C_{ij}\right]}\label{eq:IP_Jij},\\
h_{i}^{IP} & =\frac{1}{4}\sum_{j\neq i}\log\frac{\left[\left(1+m_{i}\right)\left(1+m_{j}\right)+C_{ij}\right]\left[\left(1+m_{i}\right)\left(1-m_{j}\right)-C_{ij}\right]}{\left[\left(1-m_{i}\right)\left(1+m_{j}\right)-C_{ij}\right]\left[\left(1-m_{i}\right)\left(1-m_{j}\right)+C_{ij}\right]}-\left(N-2\right)\text{atanh}m_{i}.\label{eq:IP_hi}
\end{align}

As a remark, notice that the last term in (\ref{eq:IP_hi}) avoids
the over-counting of the terms depending only on the magnetization of
site $i$, as noted in \citep{roudi_statistical_2009}.
The Independent Pair approximation can also be derived from a direct minimization of the Bethe Free energy with respect to the first and second moments \citep{SP_Ricci_Tersenghi}. \\
Now, by plugging the expression of the cavity parameters \eqref{eq:cav_field_ij}-\eqref{eq:cav_coupling_Sij} into \eqref{IP_first} we can compute the final expressions for the inferred couplings $\{J_{ij}^{*} \}$ and fields $\{h_{i}^{*}\}$ for the generic framework proposed in the previous section (so far, we have not made explicit the closure choices associated with the DC and the EP scheme, and we refer to Appendix (\ref{sec:details_computation}) for further computational details):

\begin{align}
J_{ij}^{*} & =J_{ij}^{IP}-\left(\boldsymbol{\Sigma}^{-1}\right)_{ij}-\frac{\Sigma_{ij}}{\Sigma_{ii}\Sigma_{jj}-\Sigma_{ij}^{2}}\qquad\forall i\neq j\label{eq:Jij_DCgeneral}\\
h_{i}^{*} & =h_{i}^{IP}+\left(N-2\right)\text{atanh}m_{i}-\sum_{j\neq i}\frac{\Sigma_{jj}\mu_{i}-\Sigma_{ij}\mu_{j}}{\Sigma_{ii}\Sigma_{jj}-\Sigma_{ij}^{2}}+\left(\boldsymbol{\Sigma}^{-1}\boldsymbol{\mu}\right)_{i}\qquad\forall i.\label{eq:hi_DCgeneral}
\end{align}
To get an explicit expression in terms of first and second empirical moments $\left(\boldsymbol{m},\boldsymbol{C}\right)$, it is necessary to fix a set of closure equations. In particular, by using respectively \eqref{DCclosure_all} for the DC closure and \eqref{EPclosure_all} for the EP closure, at fixed magnetizations and correlations we get
\begin{equation}
\mu_{i}^{DC}=m_{i}\quad\forall i,\qquad\Sigma_{ii}^{DC}=\frac{m_{i}}{\text{atanh}m_{i}}\quad\forall i,\qquad\Sigma_{ij}^{DC}=C_{ij}\sqrt{\frac{\Sigma_{ii}^{DC}}{C_{ii}}\frac{\Sigma_{jj}^{DC}}{C_{jj}}}\quad\forall i\neq j\label{eq:DC_closure_fixedmoments}
\end{equation}
\begin{equation}
\mu_{i}^{EP}=m_{i}\quad\forall i,\qquad\Sigma_{ii}^{EP}=1-m_{i}^{2}\quad\forall i,\qquad\Sigma_{ij}^{EP}=C_{ij}\quad\forall i\neq j.\label{eq:EP_closure_fixedmoments}
\end{equation}

Finally, by inserting (\ref{eq:DC_closure_fixedmoments}) or (\ref{eq:EP_closure_fixedmoments})
into \eqref{eq:Jij_DCgeneral}, \eqref{eq:hi_DCgeneral} we get the
final expression for the inferred parameters under DC / EP approximations.
We report in the next sub-sections the final expressions.

\subsubsection{EP}
\begin{align}
J_{ij}^{EP} & =J_{ij}^{IP}-\left(\boldsymbol{C}^{-1}\right)_{ij}-\frac{C_{ij}}{C_{ii}C_{jj}-C_{ij}^{2}}\quad\forall i\ne j\label{eq:Jij_EP}\\
h_{i}^{EP} & =h_{i}^{IP}+\left(N-2\right)\text{atanh}m_{i}-\sum_{j\neq i}\frac{C_{jj}m_{i}-C_{ij}m_{j}}{C_{ii}C_{jj}-C_{ij}^{2}}+\left(\boldsymbol{C}^{-1}\boldsymbol{m}\right)_{i}\qquad\forall i\label{eq:hi_EP}
\end{align}
where $C_{ii}=1-m_{i}^{2}$ is the variance of spin $i$.

\subsubsection{DC}
\begin{align}
J_{ij}^{DC} & =J_{ij}^{IP}-\left(\boldsymbol{\Sigma}^{-1}\right)_{ij}-\frac{C_{ij}}{C_{ii}C_{jj}-C_{ij}^{2}}\sqrt{\frac{C_{ii}}{\Sigma_{ii}}\frac{C_{jj}}{\Sigma_{jj}}}\quad\forall i\ne j\label{eq:Jij_DC}\\
h_{i}^{DC} & =h_{i}^{IP}+\left(N-2\right)\text{atanh}m_{i}-\sum_{j\neq i}\frac{\Sigma_{jj}m_{i}-\Sigma_{ij}m_{j}}{\Sigma_{ii}\Sigma_{jj}-\Sigma_{ij}^{2}}+\left(\boldsymbol{\Sigma}^{-1}\boldsymbol{m}\right)_{i}\qquad\forall i\label{eq:hi_DC}
\end{align}
where $\Sigma_{ii}$,$\Sigma_{ij}$ are defined in (\ref{eq:DC_closure_fixedmoments}).

Eqs. \eqref{eq:Jij_EP}-\eqref{eq:hi_EP} and \eqref{eq:Jij_EP}-\eqref{eq:hi_EP} require only the evaluation of the first two moments of the empirical distribution,  and their computational cost is dominated by a (single) matrix inversion, that scales as $O(N^3)$ (by using standard Gauss-Jordan decomposition).

\subsection{Relation with Sessak-Monasson approximation}

Perhaps surprisingly, the expression of the couplings with EP closure (\ref{eq:Jij_EP})
has been already derived in literature by Sessak and Monasson in \citep{sessak_small-correlation_2009}. The derivation substantially differs from the one proposed here: the authors performed a small correlations expansion of the Ising free energy at fixed magnetization and $2$-points correlations, easily represented diagrammatically. More precisely, they present an analytic summation of all the loops and $2$-spin diagrams that leads to an expression for the inferred couplings exactly equal to (\ref{eq:Jij_EP}). As noticed in \citep{roudi_ising_2009}, the expression can be viewed as a combination of the Mean Field approximation (i.e. the term $-\boldsymbol{C}^{-1}$), the Independent Pair approximation, plus an additional term that cures the over-counting of 2-spin diagrams in the first two terms. 

Conversely, the expression of the inferred field in \citep{sessak_small-correlation_2009} does not allow for these diagrams to be summed up and it is presented as a series expansion, up to the fourth order in $\beta$ \footnote{In practice, the small correlation expansion can be easily performed by putting $C_{ij}\to\beta C_{ij}$, where $\beta$ plays the role of a fictitious inverse temperature, and then expanding over $\beta$. In this notation, the $0$-th order term corresponds to the non-interacting model, that can be easily handled.}.

Interestingly, the analogy with EP closure suggests that also the expression $h_{i}^{EP}$
should be similar, and indeed we found that (\ref{eq:hi_EP}) gives
the exact series expansion presented in \citep{sessak_small-correlation_2009},
apart from a $0$-th order term. Therefore, the expression of inferred fields under the Sessak-Monasson (SM) approximation, obtained by summing loop and 2-spin diagrams, can be re-phrased as:
\begin{equation}
h_{i}^{SM}=h_{i}^{EP}+\left(N-2\right)\left[\frac{m_{i}}{1-m_{i}^{2}}-\text{atanh}m_{i}\right]+O\left(\beta^{4}\right),\label{eq:hi_SM_vsEP}
\end{equation}
where $h_{i}^{EP}$ is defined in (\ref{eq:hi_EP}). Eventually, further investigation would be needed to verify the correctness of (\ref{eq:hi_SM_vsEP}) at higher orders in the series expansion; we leave this point for future works.

DC solution for the IIP (\ref{eq:Jij_DC})-(\ref{eq:hi_DC}) has a similar structure, but some differences arise from the different closure used. As a result, expanding (\ref{eq:Jij_DC}) and (\ref{eq:hi_DC}) leads to the same diagrams reported in \citep{sessak_small-correlation_2009}: the functional dependency on the correlations is the same (which is obvious since \eqref{eq:DC_closure_fixedmoments} is a linear function of correlations) but the magnetizations-dependent polynomial coefficients in front of each term are different. Interestingly, in the limit of zero magnetizations, $\boldsymbol{\Sigma}^{DC}\to\boldsymbol{C}$ and the expression of
the inferred couplings tends to the Sessak-Monasson approximation, namely
$J_{ij}^{DC}\to J_{ij}^{EP}=J_{ij}^{SM}$.

\section{Results\label{sec:Results}}

We performed extensive and controlled numerical simulations on synthetic graphs, by varying the graph topology and model settings, i.e. the distribution of the couplings and external fields. We first compare our performances to those of the SM and EP approximations in order to establish whether the DC-based reconstruction carries some improvement with respect to these two similar strategies. To this purpose, we show the results obtained on a fully connected frustrated model of $20$ nodes where the exact computation of the trace governing the calculation of $Z$, and therefore the equilibrium statistics of the model, can be performed in a reasonable time. 
Secondly, we compare DC-based inference to other state-of-the-art techniques for this task, on synthetically generated and relatively small instances, where the partition function can again be computed in feasible time. Thus the magnetizations and the covariance matrices used for solving the inverse problems are exact.
Finally, we face the comparison to Pseudo-Likelihood (PL) maximization \citep{Besag_PL,Aurell_Ekeberg_PL} which is the out-performer for the IIP as shown in \citep{nguyen_inverse_2017}. For this last experiment, we sample from the true distribution a reasonable number of configurations used both to run PL and to compute the empirical statistics needed from the other inference techniques. We remark that this last method uses a different type of data, that is the raw empirical configurations, at difference with all the other techniques which are derived from a limited set of statistical observables, i.e. the empirical first and second moments.
We also show here how the inference performances are affected by the number of samples; indeed, having access to a copious set of configurations improve the PL estimates and allow us to accurately compute the first and second moments (so both PL and all fixed-statistics methods will benefit from it). %conversely, PL running time is severely affected by the number of samples at difference with the fixed-statistics methods whose computing time is independent of the number of configurations used to estimate the empirical statistics. \\
\\
In all the cases illustrated hereafter, we measure the quality of the inference by computing the reconstruction error of the inferred couplings and fields
w.r.t. the true ones as
\begin{equation}
\Delta_{J}=\sqrt{\frac{\sum_{i<j}\left(\beta J_{ij}^{t}-J_{ij}\right)^{2}}{\sum_{i<j}\left(\beta J_{ij}^{t}\right)^{2}},}\qquad\Delta_{h}=\sqrt{\frac{\sum_{i}\left(\beta h_{i}^{t}-h_{i}\right)^{2}}{\sum_{i}\left(\beta h_{i}^{t}\right)^{2}},}\label{eq:errors_Jandh}
\end{equation}
where $\left(\beta h_{i}^{t},\beta J_{ij}^{t}\right)$ are the true
model parameters. Another important measure of the inference quality that will be used in the last paragraph is the area under the Receiver Operating Characteristic (ROC) curve, denoted as AUC. The latter quantifies how good the detection of present/absent couplings is, independently on their strength \citep{FAWCETTROC}. For each scenario, simulations were run on $n$
different instances by varying the seed of the random number generator used to generate the topology and model parameters. Therefore, all the following plots
show the average and standard error for the three measures (Eq. \eqref{eq:errors_Jandh} and AUC):
\begin{equation}
\bar{f}=\frac{1}{n}\sum_{\alpha=1}^{n}f^{\alpha}\qquad\delta f=\frac{1}{\sqrt{n}}\sqrt{\frac{1}{n-1}\sum_{\alpha=1}^{n}\left(f^{\alpha}-\bar{f}\right)^{2}}\label{eq:mean_and_std_errors}
\end{equation}
where $f\in\left\{ \Delta_{J},\Delta_{h}, AUC \right\} $ and the summation runs over the different instances. From a technical point of view, a small diagonal regularization $\varepsilon=10^{-10}$ has been added to the empirical/exact covariance matrix $\boldsymbol{C}$ to prevent numerical problems in the inversion of this matrix, arising especially at low temperatures. \\

\subsection{Comparison to Sessak-Monasson approximation}
\begin{figure}
\includegraphics[width=0.49\textwidth]{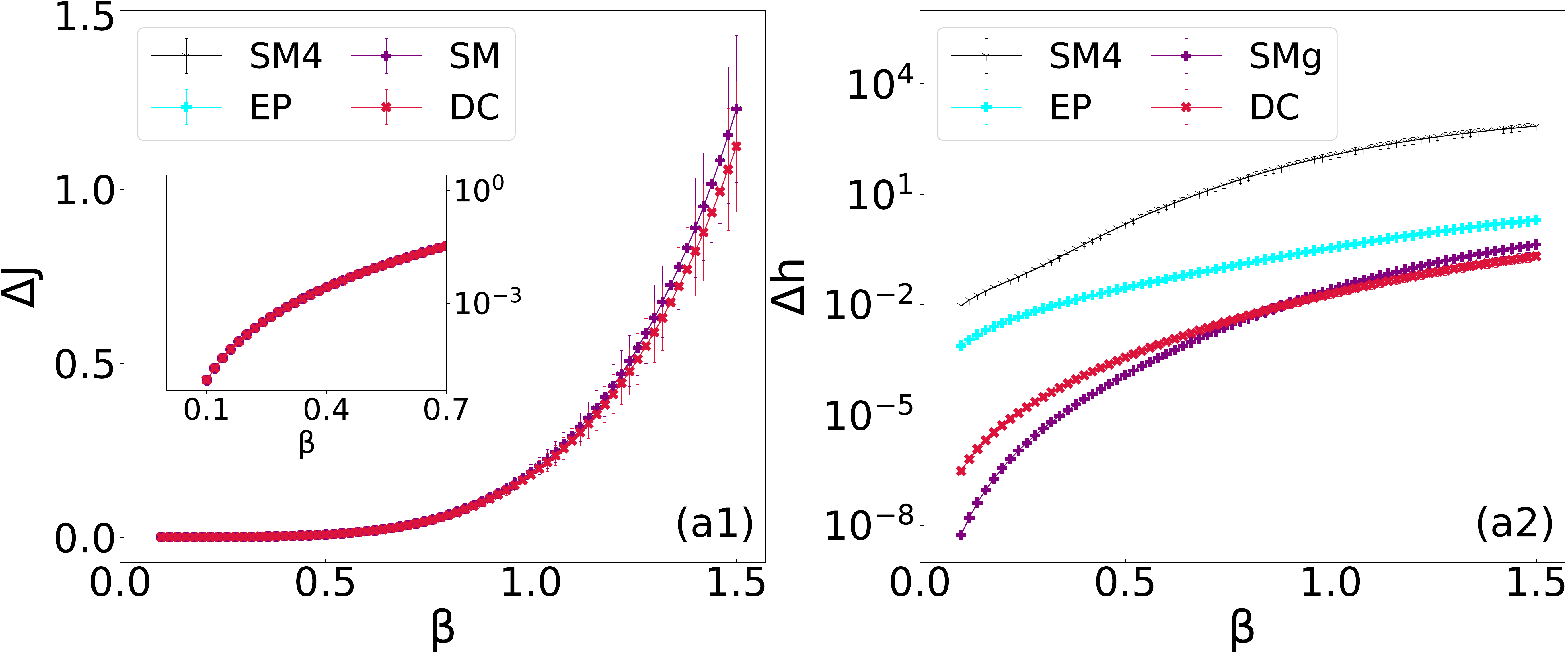}\hfill
\includegraphics[width=0.49\textwidth]{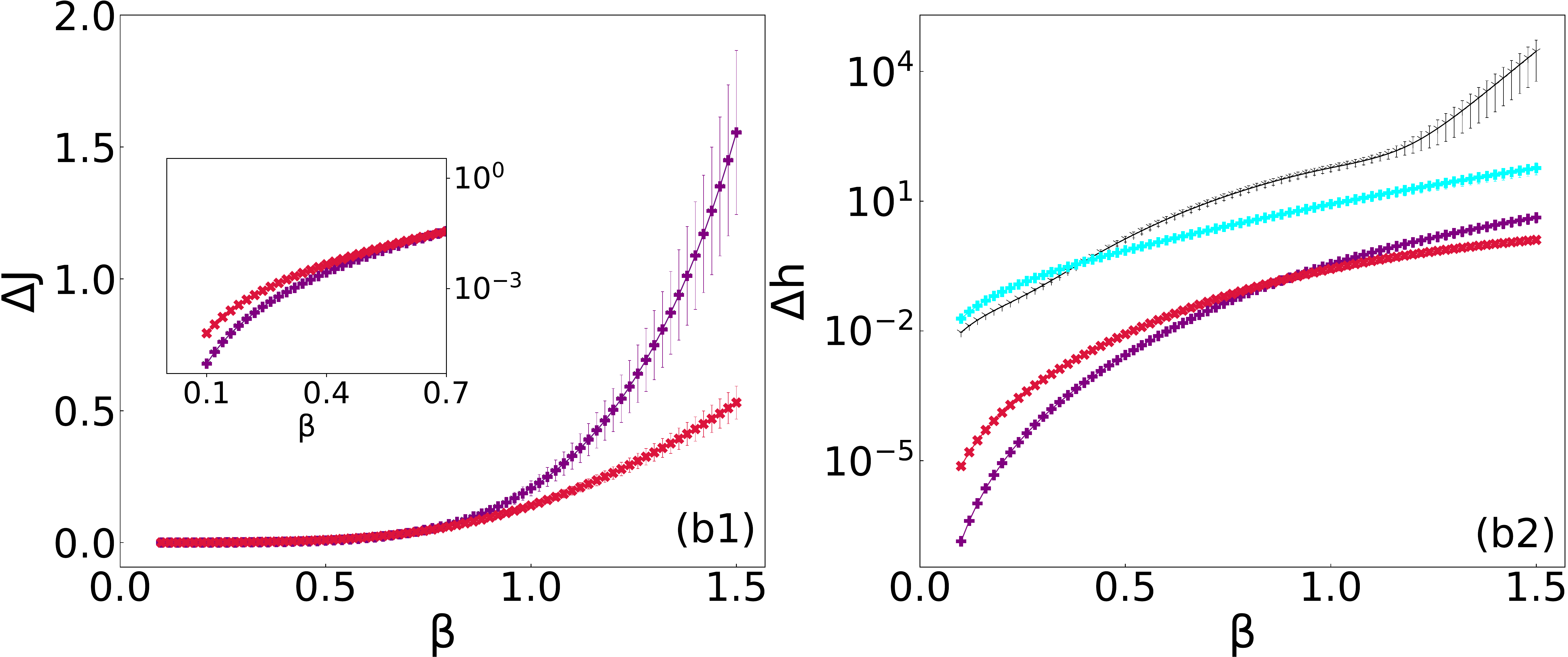}

\caption{Reconstruction on a fully connected graph of $N=20$ nodes. The binary couplings satisfy $J_{ij}\sim\pm\frac{\beta}{\sqrt{N}}$ and the fields are uniformly distributed, $h_{i}\sim h_{0}U\left(0,\beta \right)$. 
(a): $h_{0}=0.1$; (b): $h_{0}=0.5$. For each scenario, we show the error w.r.t the couplings $\Delta_{J}$ and the fields $\Delta_{h}$, averaged on $30$ instances. The insets show the log-scale behaviour at high temperatures (low $\beta$).\label{fig:SK_20}}
\end{figure}

We start from a comparison between SM, EP and DC on a fully connected frustrated model in the presence of random external fields. As discussed in the previous section, the EP and SM inference of couplings coincides and it is given by Eq. \eqref{eq:Jij_EP}; for what concerns the fields' reconstruction, we compared the EP and DC expressions (resp. Eq. \eqref{eq:hi_EP} and Eq. \eqref{eq:hi_DC}), to the small correlation expansion up to $4-$th order in $\beta$ presented in \citep{sessak_small-correlation_2009} (labelled as \textit{SM4}) and our closed-form expression given by Eq. \eqref{eq:hi_SM_vsEP} (labelled as \textit{SMg}). We show in Fig. \ref{fig:SK_20} the reconstruction performances for a fully connected graph, of $N=20$ spins. Here the distribution of the couplings is binary, that is $J_{ij} \sim \pm\frac{\beta}{\sqrt{N}}$ while the external fields are distributed uniformly, namely $h_{i} \sim \beta h_{0} U(0,1)$, for a certain value of the scale parameter $h_{0}$. The small size of the system allows us to enumerate all the $2^N$ equilibrium configurations and thus to exactly compute the magnetizations and the covariance matrix of the model. \\
We show in the two plots on the left (on the right) in Fig. \ref{fig:SK_20} the error reconstruction of the couplings and the fields when $h_{0} = 0.1$ ($h_{0} = 0.5$). The behaviour of DC and SM, on $\Delta_{J}$, is similar at small field scale $h_{0}$ (Fig \ref{fig:SK_20}, panel a) because the magnetizations are relatively small and therefore DC estimate tends to SM. However, as we can notice from the sub-plots on the right of Fig. \ref{fig:SK_20} (panel b), when the scale parameter associated with the fields increases, DC reconstruction improves the estimate of the model parameters, if compared to SM. Interestingly, the gap between the two methods increases as the temperature decreases suggesting the DC reconstruction has to be preferred when dealing with this kind of data. Notice also that the standard error of DC reconstruction remains small also at low temperatures, while the SM one increases. For what concerns the fields reconstruction, the inference under EP and SM truncated to the fourth order is quite poor, especially at low temperatures and in presence of strong external fields. On the other hand, the closed form expression given by \eqref{eq:hi_SM_vsEP} for the SM reconstruction gives a good estimate of the fields at high temperatures, while at low temperatures, again, DC outperforms SM. We retrieve qualitatively the same results under different conditions, that is when the model parameters are distributed according to different probabilities, i.e. Gaussian fields and/or couplings. 

\subsection{Reconstruction using exact statistics}
\label{subsec:rec_exact}
In this section we show results on different sparse topologies, i.e. graphs with low average connectivity, by comparing DC reconstruction to other mean-field like methods for the IIP: the Independent Pair (IP) approximation \eqref{eq:IP_Jij}-\eqref{eq:IP_hi}, Sessak-Monasson (SM) approximation, Mean-Field (MF), TAP equations and Susceptibility Propagation (SP). In particular, for what concerns SP, we implemented the analytic solution derived in \citep{SP_Ricci_Tersenghi} that avoids to run the iterative algorithm introduced in \citep{SPMezardMora}. For the last three methods, Mean-Field, TAP and SP, we report in the Supplementary material the final expression
of the inferred couplings and fields, and we refer to \citep{nguyen_inverse_2017,SP_Ricci_Tersenghi} for an extensive discussion about of them. Moreover, from now on, Eq. \eqref{eq:hi_SM_vsEP} is used to reconstruct the external fields under the SM approximation, as the $4-$th truncated expansion typically gives quite large errors and  because of this it will not be shown.
We remark that all the above mean-field techniques give a closed-form solution for the inferred parameters: with the exception of IP, all the others require a (single) matrix inversion, just as Density Consistency.
We performed simulations for many combinations of the couplings and fields distributions (Gaussian $\mathcal{N}(0,1)$, uniform positive $U(0,1)$, binary $\pm1$ and constant) and graph topology: we used Erd\H{o}s-R\'{e}nyi and random regular graphs with different mean (resp. fixed) connectivity, scale-free graphs generated by the preferential attachment, i.e. using the Barabasi-Albert model \citep{BarabasiAlbert}, regular lattices (in particular, diluted triangular and square lattices where only a fraction $p \in [0,1] $, called dilution coefficient, of the set of links of the full structure, is considered). 

\begin{figure}
\includegraphics[width=0.49\textwidth]{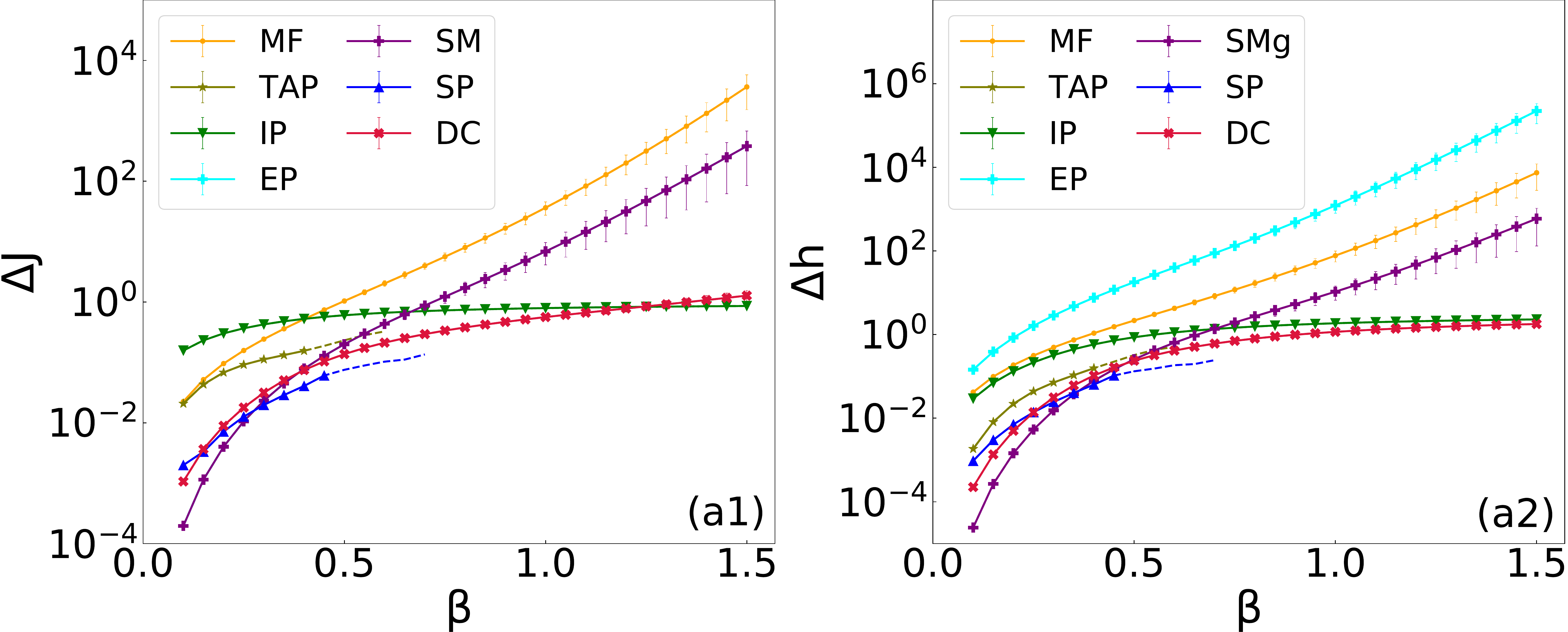}\hfill
\includegraphics[width=0.49\textwidth]{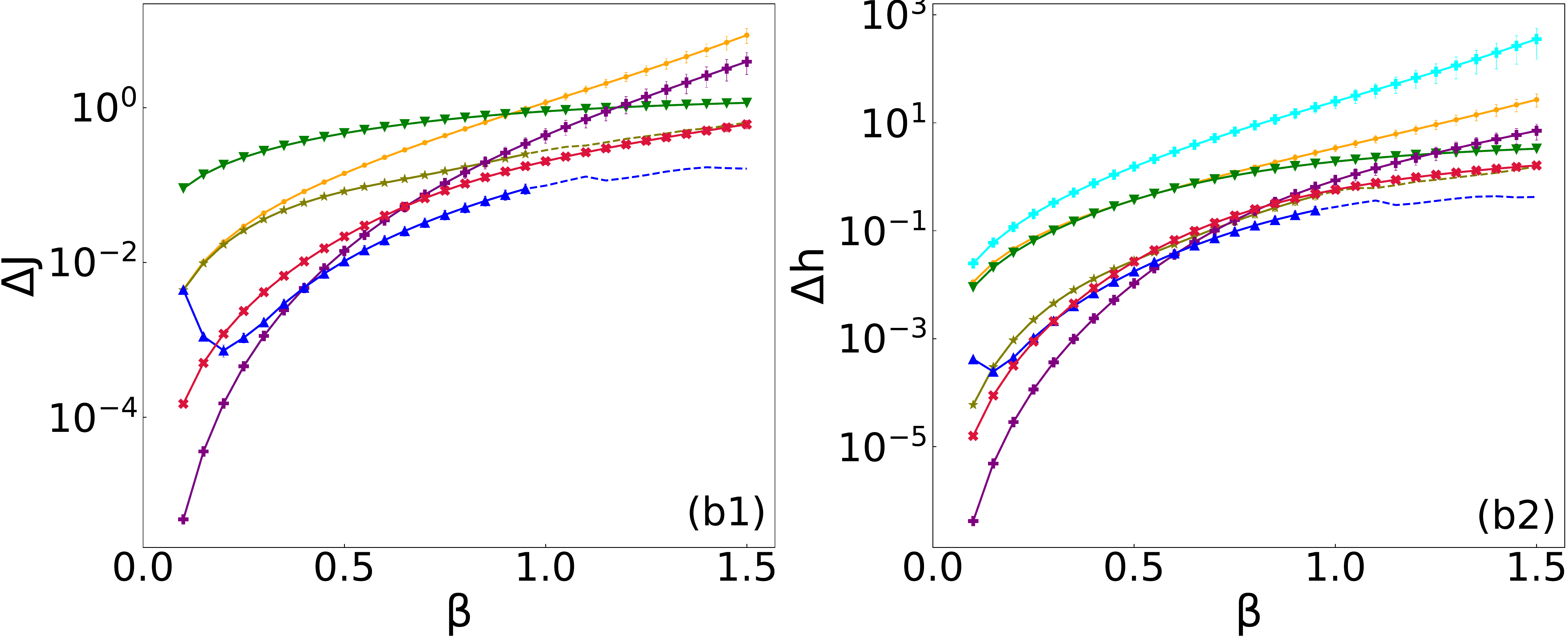}\\
\vspace{.5cm}
\includegraphics[width=0.49\textwidth]{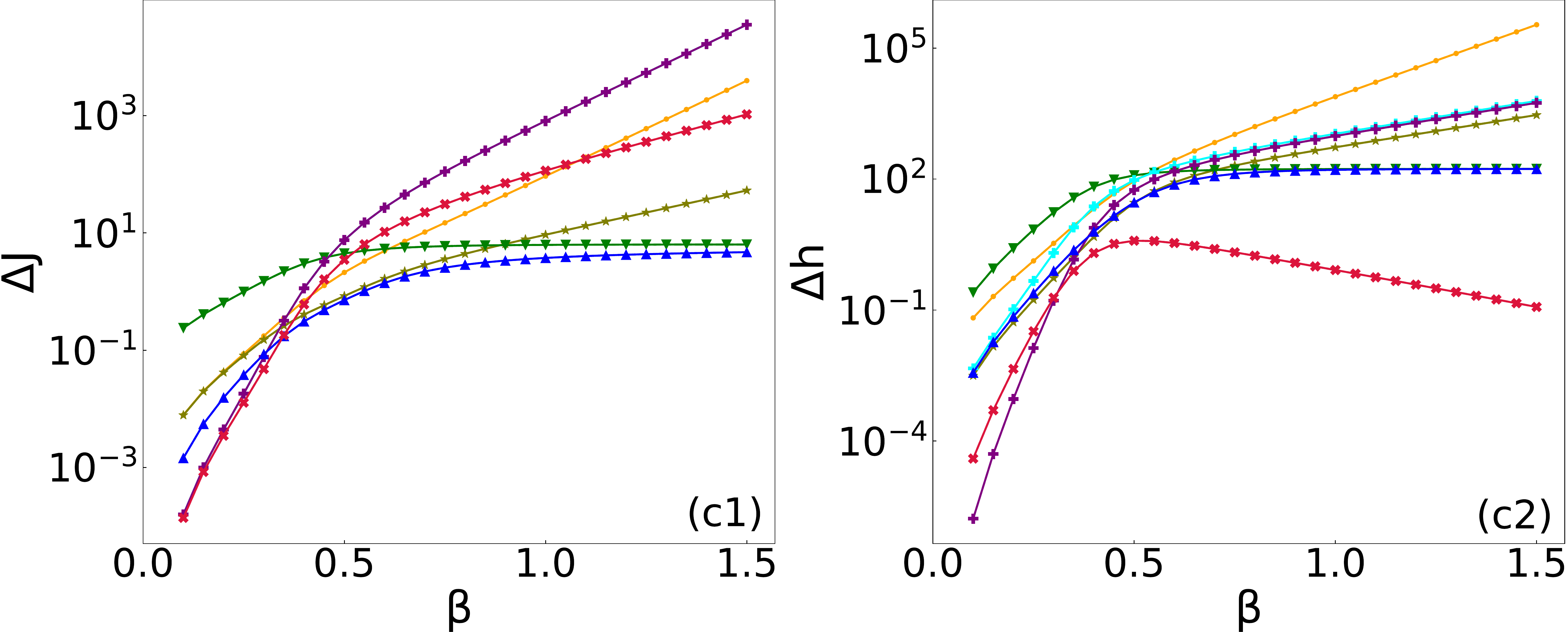}\hfill
\includegraphics[width=0.49\textwidth]{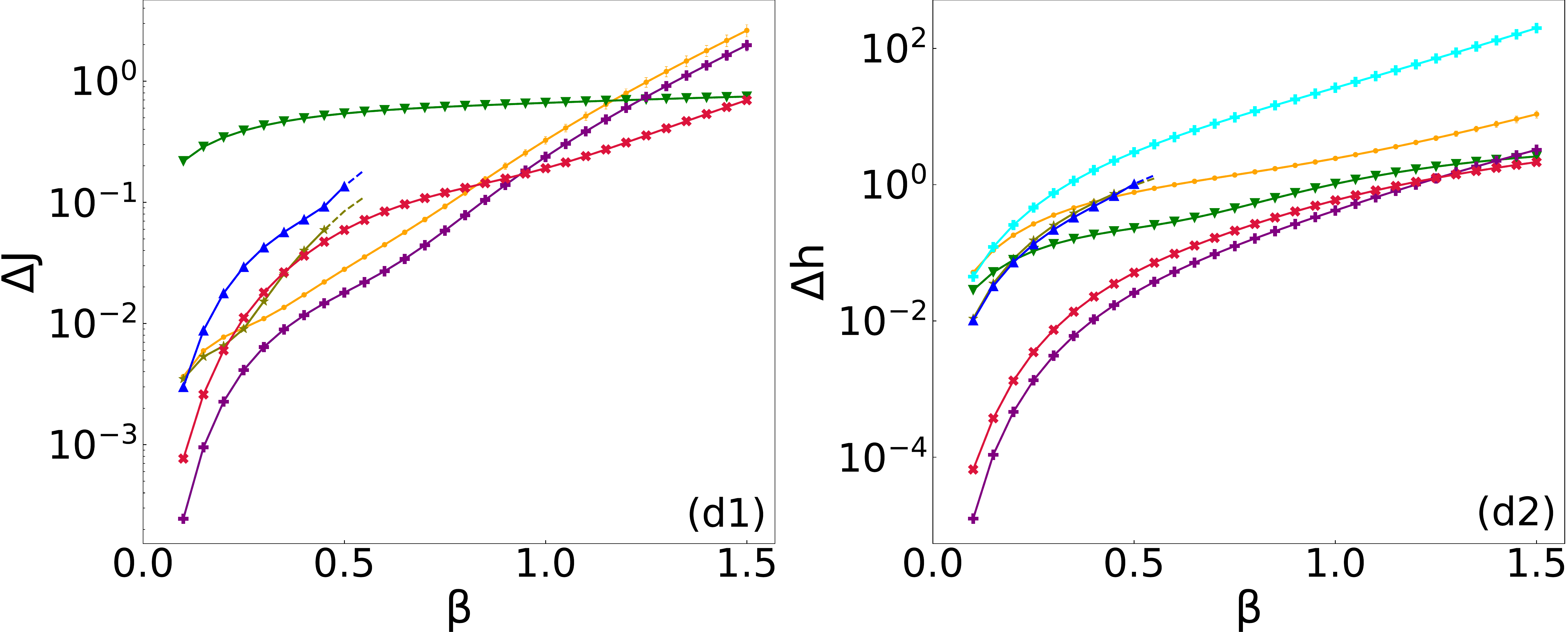}
\caption{Reconstruction in sparse topologies from exact observables. (a) Erd\H{o}s-R\'{e}ny graph with $N=20$ and mean connectivity $k=3$, Gaussian couplings $J_{ij}\sim  N(0,\beta^{2})$, constant fields $h_{i} = \beta$. (b) $2D$ square lattice with periodic boundary conditions of $N=4^2$ nodes and dilution coefficient $p=0.7$, uniform positive couplings $J_{ij}\sim  U(0,\beta)$, binary fields $h_{i} \sim \pm 0.5 \beta $. (c) Random regular graph with $N=20$ and fixed connectivity $k=4$, constant couplings $J_{ij} = \beta$, constant fields $h_{i} = 0.1 \beta$. (d) Triangular lattice with $N=4^2$ anti-ferromagnetic couplings $J_{ij} = -\beta$, binary fields $h_{i} = \pm 0.6 \beta$. All the plots show respectively the error over couplings and over fields, averaged on $n=50$ instances. SP and TAP reconstructions are shown as a dashed line where the number of instance giving physical solution $n^{*}$ is $n\slash 2 < n^{*} < n$ larger than half of them, and it is not shown for $n^{*} < n\slash 2$. \label{fig:Trace_all}}
\end{figure}

A selected subset of results is shown in Fig. \ref{fig:Trace_all}. We can identify a general behaviour of the different methods: SM is always better than MF but both of them give large errors at low temperatures. TAP and SP outperform MF but suffer from numerical problems when the scale parameter for the fields is large and the temperature of the system is low (see for instance the results for the Erd\H{o}s-R\'{e}nyi and graph the triangular lattice in Fig. \ref{fig:Trace_all} (a) and (d) respectively). Here both TAP and SP equations have no fixed points in these regimes as the arguments of the square roots appearing in the expressions of the couplings become negative, as noted in \citep{SP_Ricci_Tersenghi}.
In particular, in Fig. \ref{fig:Trace_all}, we differentiate the regimes in which TAP and SP provide physical solutions in all the $n$ instances, by using both dots and lines, from those in which at least one time they do not find a solution (here we plot lines, without dots). When these two methods are not able to give a physical solution to more than half of the instances (here more than 25 out of 50),  we do not show any line. \\
DC turns out to significantly outperform SM in almost all cases at low temperatures and it provides comparable estimates to SP at small $\beta$. A different behaviour can be noted in Fig. \ref{fig:Trace_all} (c), where DC performs worse than other methods for the couplings reconstruction, but it gives a very good estimation for the fields.
%Qualitatively, it seems that imposing the DC closure in the computation of the covariance matrix, denoted here as $\Sigma^{DC}$, cures the numerical instabilities revealed by TAP and SP.
%\textcolor{red}{Giova: forse metteri che cura le numerical instabilities di SM no? perché alla fine TAP e SP non convergono ma per altri motivi} \textcolor{blue}{le radici negative non le hanno SP eTAP? è la chiusura che permette di avere una matrice di cov invertibile, e poi di fare l'inferenza. non scriviamo che SM ha instabilità numeriche ma che l'approssimazione di DC migliora quella di SM}\textcolor{red}{Giova: si, però quello che direi io è che DC migliora le instabilità numeriche che sono proprie di SM grazie alla chiusura che imponiamo, per TAP e SP il problema c'è ma è di tipo diverso direi} \textcolor{blue}{Allora forse è meglio togliere proprio questa frase...}

\subsection{Inference using sampled configurations}

To compare the performances of DC against the reconstruction performed by PL, we needed to first sample a set of $M$ configurations from the true model.
The Monte Carlo Markov Chain (MCMC) algorithm used for this task is the Gibbs sampler \citep{Geman84}. Starting from a uniformly random configurations of $N$ spins, to fairly sample independent configurations, we first let the MCMC dynamics equilibrate by performing $M$ steps. Then, we run the same dynamics for other $Md$ steps and we collect $M$ samples ($1$ every $d$).  We remark that each ``step'' here corresponds to $N$ sequential Gibbs-sampling attempted flips, one for each spin performed following a random permutation of the indices. In all the simulations, $M=10^{5}\text{, }d=10^{2}$. 
%In these runs, we considered larger systems of $N \sim 50$ spins interacting through the same topologies illustrated in Section \ref{subsec:rec_exact}. 

\begin{figure}
\includegraphics[width=0.49\textwidth]{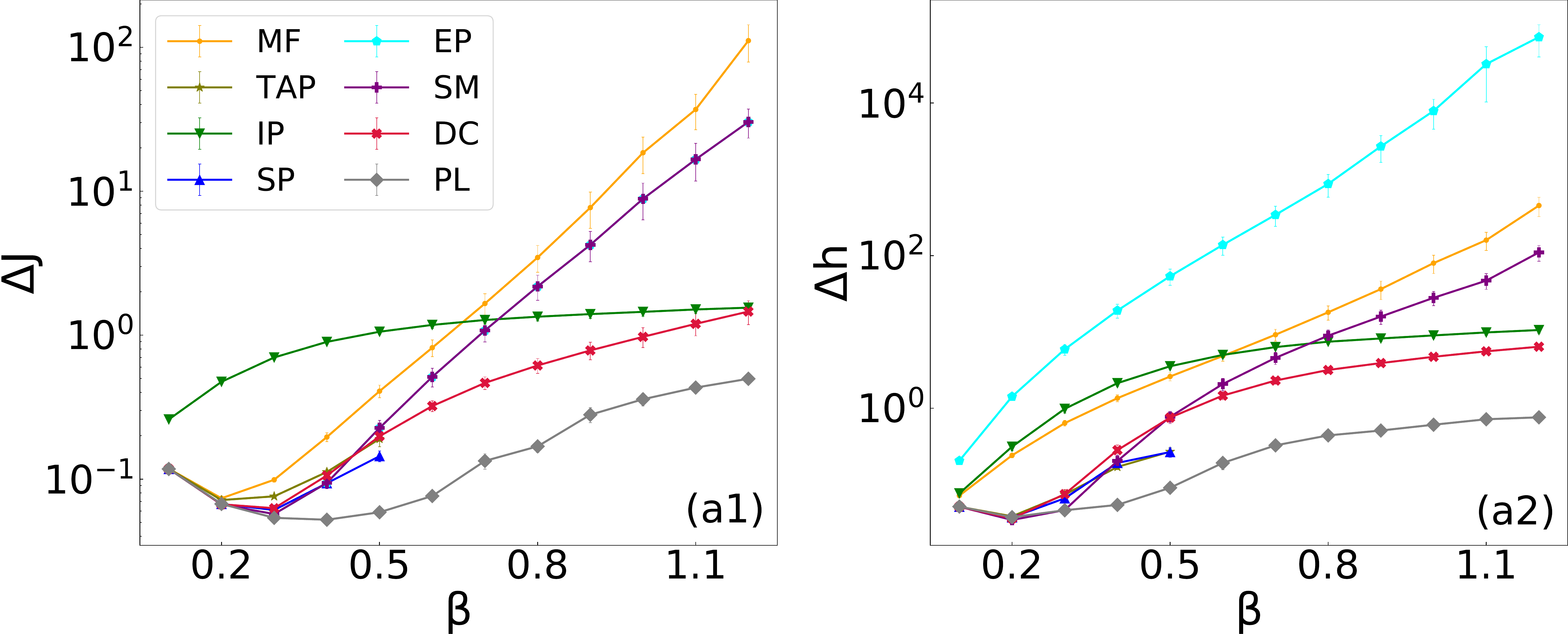}\hfill
\includegraphics[width=0.49\textwidth]{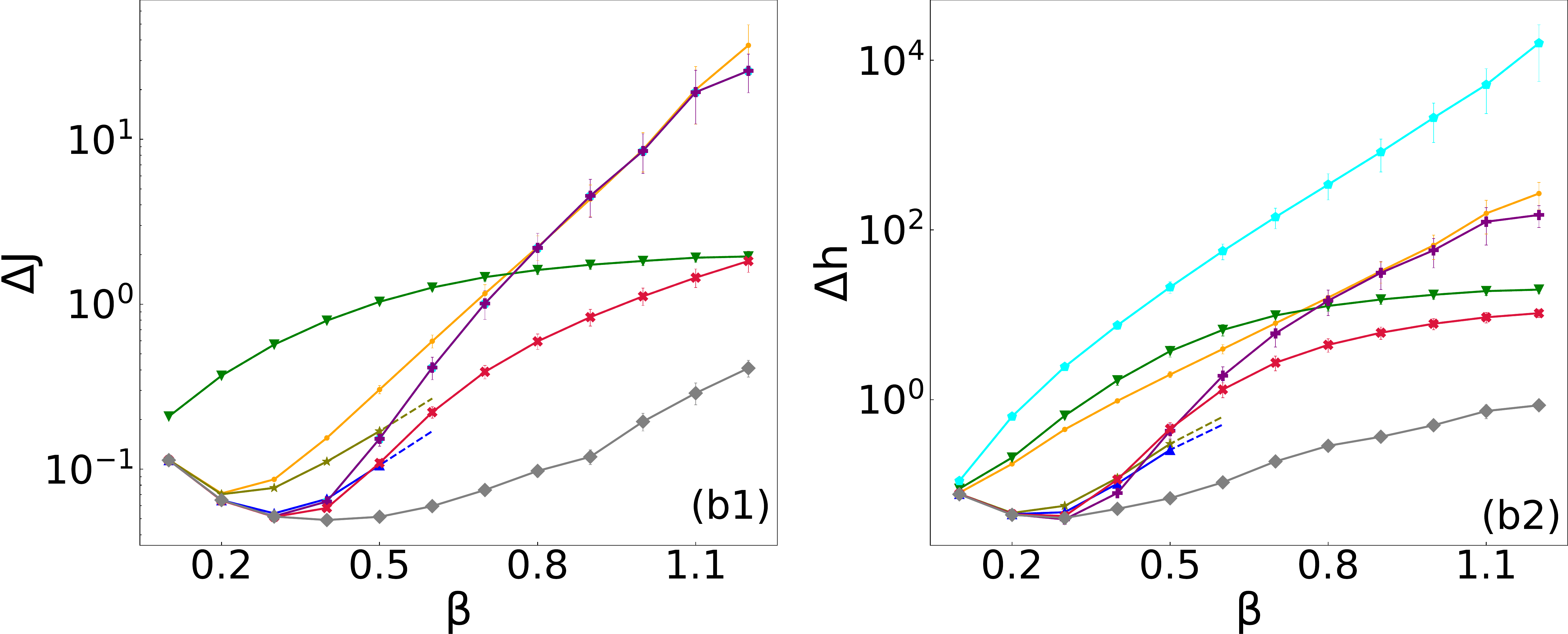}\\\includegraphics[width=0.49\textwidth]{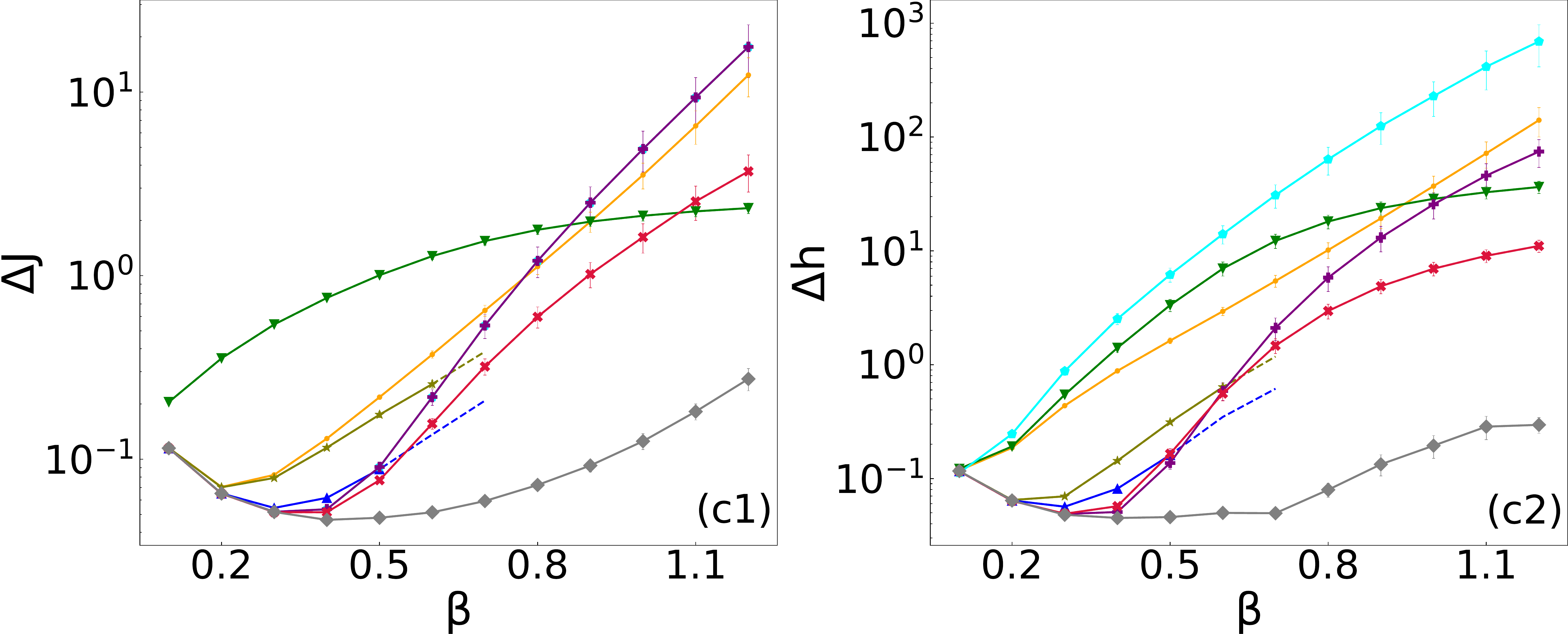}\hfill
\includegraphics[width=0.49\textwidth]{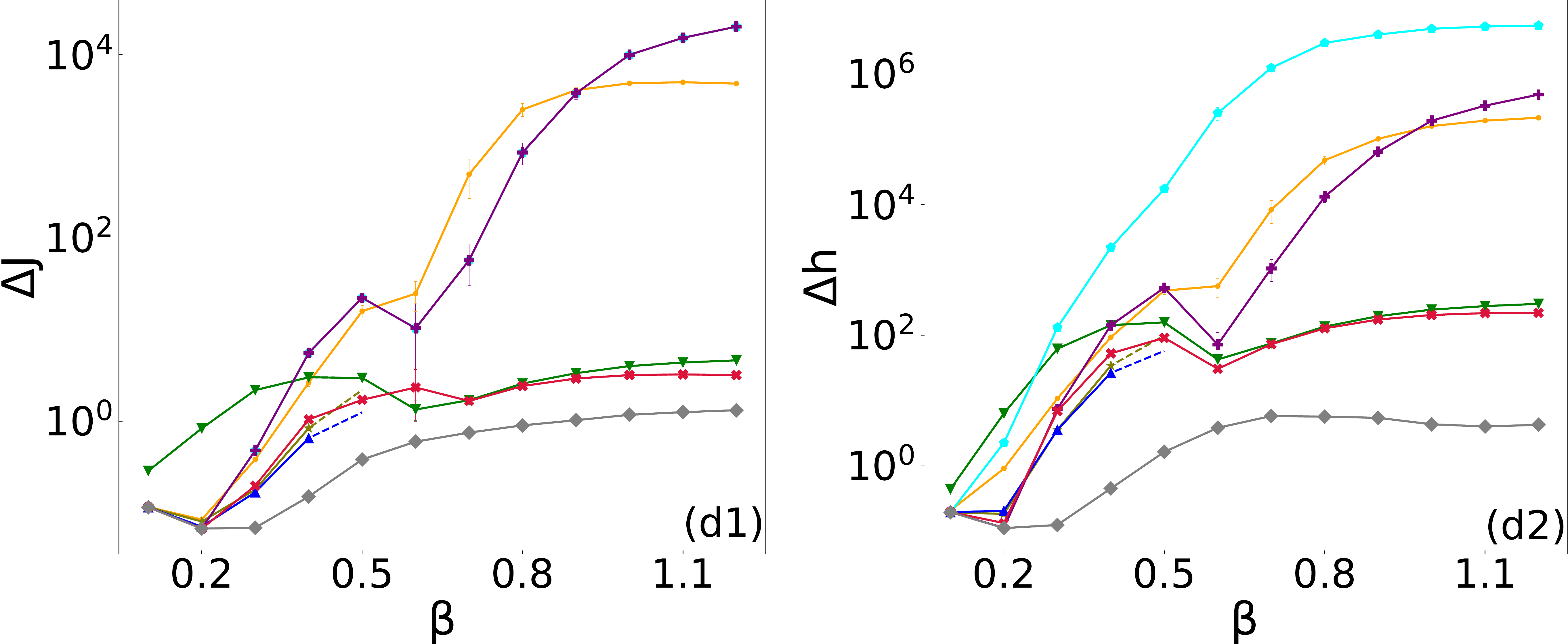}

\caption{Reconstruction in sparse topologies using sample averages. (a) Barabasi-Albert \cite{BarabasiAlbert} graph with $N=50$, $n_{0}=k=2$, binary couplings $J_{ij}\sim \pm \beta$ and constant fields $h_{i}=0.7\beta$. (b) $2d$ square lattice with $N=7^2$ and PBC, binary couplings $J_{ij}\sim \pm \beta$ and constant fields $h_{i}=0.5\beta$. (c) Random regular graph with $N=50$ and fixed connectivity $k=4$, binary couplings $J_{ij} \sim \pm \beta$, binary fields $h_{i} \sim \pm 0.3 \beta$. d) Erd\H{o}s-R\'{e}ny graph with $N=50$ and mean connectivity $k=4$, constant couplings $J_{ij} = \beta$, uniform fields $h_{i} \sim 0.3 U(0,\beta) $. All the plots show respectively the (log-scale) error over couplings and over fields, averaged on $20$ instances. \label{fig:Sampling_all}}

\end{figure}

%with a small regularization term to avoid the appearance of huge fields and couplings associated with very small first and second moments, computed from a poor statistics. For similar reasons and 
To avoid numerical instabilities, the estimation of the first and the second moments, used within all the other mean-field like methods, is slightly modified by the use of a small pseudo-count $\lambda=\frac{1}{M}$ \citep{barton_cocco_monasson_pseudocount}. Generally, the effect of the pseudo-count is to modify the distribution of a set of $N$ binary variables (in this case the empirical distribution) as a mixture between the starting one and a uniform probability density in the range $\left\{ -1,1\right\} ^{N}$. As a consequence, adding a pseudo-count $\lambda\in\left[0,1\right]$ modifies the non-connected moments as $\langle x_{i}\rangle \rightarrow (1-\lambda) \langle x_{i} \rangle$ and $\langle x_{i} x_{j}\rangle \rightarrow (1-\lambda) \langle x_{i} x_{j}\rangle$ for $i = 1,...,N, j\neq i$. Finally, PL maximization is performed up to a numerical precision of $10^{-4}$ with no regularization term; we used the implementation provided in \citep{PlmIsing}.\\
%$\langle x_{1}^{p_{1}}\ldots x_{N}^{p_{N}}\rangle\to\left(1-\lambda\right)\langle x_{1}^{p_{1}}\ldots x_{N}^{p_{N}}\rangle,$
%each $p_{i}$ being a non-negative integer (with $\sum_{i}p_{i}>0$). 
We show in Figure \ref{fig:Sampling_all} the reconstruction error of the couplings and the fields for MF, TAP, SP, IP, SM, PL and DC for different graph topologies and different model parameters (see the details in the caption of Figure \ref{fig:Sampling_all}). The behaviour is qualitatively similar to the results shown in Fig. \ref{fig:Trace_all}. In particular, SM/EP and MF typically have a very large reconstruction error at low temperatures; on the other hand, SP and TAP inference fails to find physical solutions for the couplings and fields at low temperatures. PL seems to outperform all the methods in all the regimes considered here while DC often provides the best estimates among all the methods that use only the information about first and second moments. 
%It is also worth noting the behavior of DC at low temperature in the regimes described in panels (b) and (d) of Figure \ref{fig:Sampling_all}: here the reconstruction of the couplings is very close to that of PL and, perhaps surprisingly, the performance seems to improve as the temperature decreases. 
%We remark that also Pseudo-Likelihood suffers from convergence issues in some of the regimes analyzed: in particular, in Fig. \ref{fig:Sampling_all}, panels b and d, PL did not converge on 2 out of the 20 instances considered, and therefore the average error is computed among the remaining 18 instances and displayed as a dashed line. In these regimes, it is probably necessary to increase the regularization prior to help PL converge.
\subsubsection{Effect of the number of samples on the performance}

\begin{figure}
\centering
\includegraphics[width=.9\textwidth]{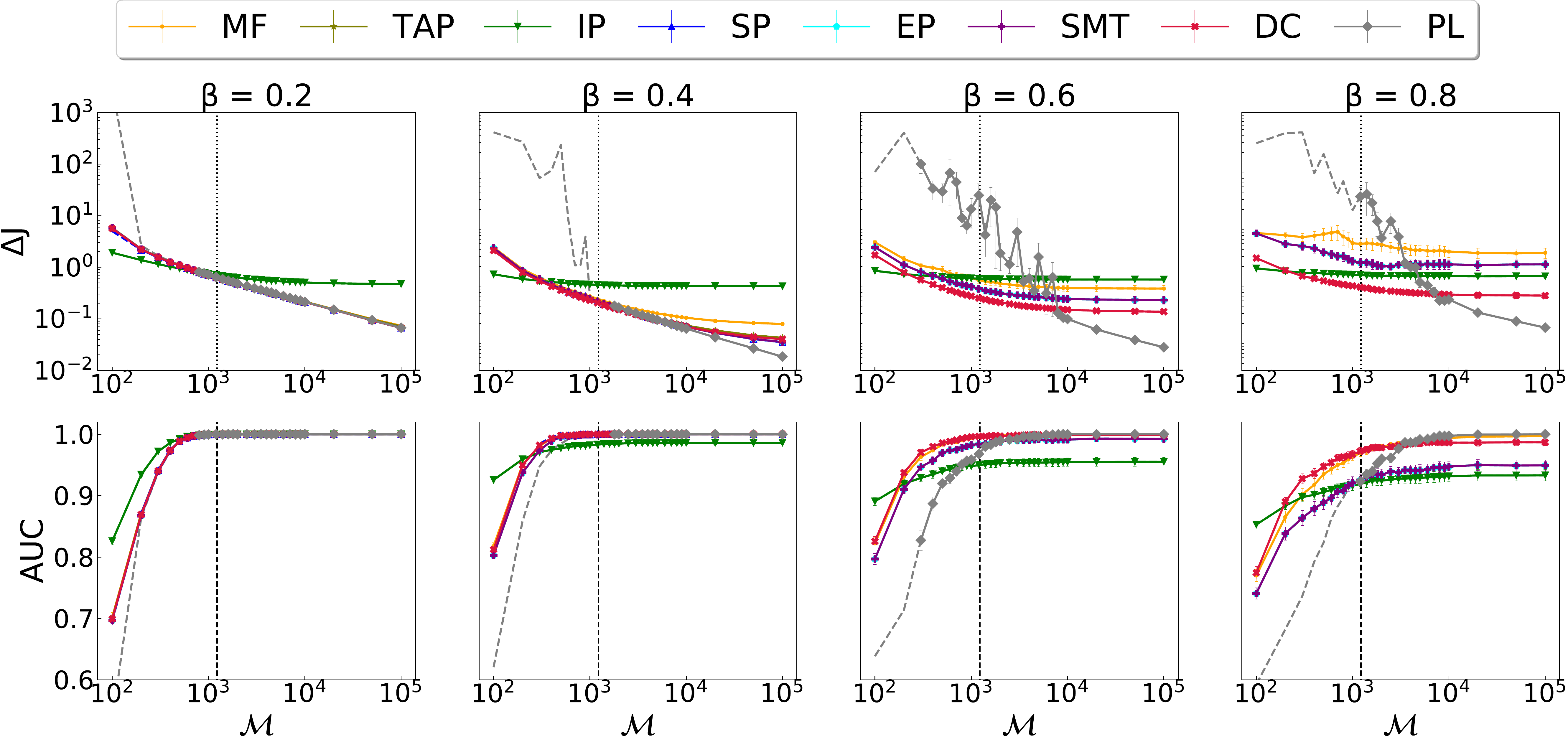}

%\includegraphics[width=.95\textwidth]{figures/vsMTriang_L_7_PBC_1_Jtype_unif_Jscale_-1.0_Jparams_0.0_1.0_htype_const_hscale_0.5}
%\caption{Effect of the number of samples on the reconstruction quality. The model is a Barabasi-Albert \cite{BarabasiAlbert} graph with $N=50$, $n_{0}=k=2$, binary couplings $J_{ij}\sim \pm \beta$ and constant fields $h_{i}=0.7\beta$, i.e the same regime of Fig. \ref{fig:Sampling_all} (a). Each plot shows the average error over the couplings for a certain value of $\mathcal{M}$ (the values are shown at the top of each subplot). \label{fig:inference_vs_M}}
\caption{Effect of the number of samples on the reconstruction quality. The model is a Barabasi-Albert \cite{BarabasiAlbert} graph with $N=50$, $n_{0}=k=2$, binary couplings $J_{ij}\sim \pm \beta$ and constant fields $h_{i}=0.7\beta$, i.e the same regime of Fig. \ref{fig:Sampling_all} (a). Each plot shows the average error over the couplings (top panel) and the AUC (bottom panel) w.r.t. the number of samples $\mathcal{M}$  for different temperatures (the values are shown at the top of each subplot). The black dotted line corresponds to $M^{*} = \binom{N}{2}$, i.e. the number of couplings to be inferred. \label{fig:inference_vs_M}}

\end{figure}

Finally, we show two results that highlight the robustness of the inference methods against the number of samples, i.e. we show how the number of samples used in computing the empirical statistics affects the performances of the methods we presented in the previous section.
In this case, we first sample the true model using the same number of configurations as in Fig. \ref{fig:Sampling_all} ($M=10^5$), and then, we select the first subset of these samples for different $\mathcal{M}<M$; therefore, the $\mathcal{M}$ samples used for the inference will be characterized by the same equilibration time $M$ and de-correlation time $d$ of the full set of samples. We show in Fig. \ref{fig:inference_vs_M} the results on the same instance chosen for Fig. \ref{fig:Sampling_all}, panel a, for $\mathcal{M} \in \left[ 10^{2}, 10^{5} \right]$. We also report as a vertical and dashed line in correspondence to $M^{*} = \frac{N(N-1)}{2}$, i.e. the full set of couplings to be inferred. This number can be consider as a natural threshold which separates a harder regime in which the number of configuration used to compute the data statistics is smaller than the number of unknown, i.e. $\mathcal{M} < M^{*}$, from the easier regime where the available data overcome $M^{*}$, that is  $\mathcal{M} > M^{*}$. 
In the large temperature regime, i.e. for $\beta = \{0.2, 0.4\}$, all methods show comparable performances with the exception of IP which provides the best estimates, in terms of both $\Delta_{J}$ and AUC, when the statistics is significantly poor, that is for $\mathcal{M} = 10^{2}$. Then the accuracy of the predictions deteriorates for large $\mathcal{M}$ values. Here, PL looses its predictive power only for $\beta=0.4$ and $\mathcal{M}<M^{*}$.
In the large $\beta$ regime, the fixed-statistics methods outperform PL in a wide range of $\mathcal{M}$-values, in some cases also in the easier regime, that is $\mathcal{M}>M^{*}$, and among them, DC seems to be preferred as suggested by both metrics, $\Delta_{J}$ and AUC. Contrarly, when the available data is abundant, PL performances are the most accurate in terms of $\Delta_{J}$ and comparable to those of DC and MF in detecting of the links of the graph, as brought in light by the AUC values.
%At extreme small values of $\mathcal{M}$ (i.e. $\mathcal{M}\sim10^2$) the Independent Pair approximation has the best performance if compared to the other fixed-statistics inference methods; this is reasonable since, unlike the other methods, it requires no matrix inversion. In a large and intermediate range of $M$, even for $M<N(N+1)/2$, and, especially for large $\beta$, DC seems to be preferred among all the methods. 

\section{Conclusions\label{sec:Conclusions}}

This work addresses the fundamental problem of finding the ML parameters of an Ising model from the magnetization vector $\mathbf{m}$ and the correlation matrix $\mathbf{C}$. We showed that, similarly to what happens with other mean-field approximations (namely IP, MF, EP, TAP, SM), the DC set of equations can be solved explicitly, giving a simple analytic expression of $\mathbf{h}, \mathbf{J}$ as a function of $\mathbf{m}, \mathbf{C}$.  We showed interesting analogies and connections between the DC, EP, SM and IP methods, allowing us in particular to propose a better candidate for the inferred fields in the SM method (in place of the fourth order expansion in $\beta$ proposed in \citep{sessak_small-correlation_2009}).

We then performed a numerical comparison of the methods' accuracies in several regimes; note that the comparison to TAP and SP was often performed for a subset of instances since their equations present singularities at low temperature. Though it was not possible to identify a clear-cut winner, the accuracy of DC was often among the best ones (sometimes the single best one). Broadly, when considering different regimes, i.e. graph topologies and model parameters, DC appears as one of the most accurate and reliable approaches in a significant range of temperatures. 
For completeness, we compared also with the PL method, which uses as input the full set of samples instead of first and second empirical moments.  Although PL comes out as one of the best methods for this problem if the samples are available, it involves an iterative optimization step which can be a drawback. In particular, we found regimes in which PL has difficulty converging, especially when the number of samples is small. In particular for a small number of samples, the mean field methods seem generally preferable to PL, with DC coming out ahead of the others.

Further research should be carried out to eventually include additional prior information about sparseness by means of a regularization prior. From an analytic point of view, the freedom in choosing the DC closure equations might allow for further improvements of the inference quality. Finally, we mention that a generalization to multi-state variables (Potts-like models) might lead to other closed form expression of the inferred parameters with potential applications in computational biology. This will be the scope of future research.

\begin{acknowledgments}
We warmly thank Andrea Pagnani for sharing with us his  Pseudo-Likelihood
maximization implementation, and Alejandro Lage Castellanos for interesting discussions.
AB, GC and APM thank Universidad de La Habana for hospitality. The authors
acknowledge funding from EU Horizon 2020 Marie Sklodowska-Curie grant
agreement No 734439 (INFERNET), the PRIN project 2015592CTH and computational
resources provided by the SmartData@PoliTO (\url{http://smartdata.polito.it}) center on Big Data and Data Science.
\end{acknowledgments}

\bibliographystyle{unsrt}

%\bibliography{InverseIsing}

\appendix

\section{DC closure equations\label{app:DCeqs}}
Density consistency is based on a set of closure equations constructed in such a way that the computation of marginals is exact on acyclic graphs, just like the Bethe Approximation. In the pairwise case discussed in the present work, for each pair of spins $(i,j)$ in the original graph (which is known in the direct problem), this property is achieved by imposing a matching of the \emph{density} values between the single-node marginals of
the tilted distribution $q^{\left(ij\right)}$, defined by Eq. (\ref{eq:qtilted_ij}) and the full Gaussian measure $q$ given by Eq. \eqref{eq:dc_fullgauss} for each of the two spins $i,j$ over which the tilted distribution is defined. This condition can be rephrased as $q^{\left(ij\right)}\left(x_{i(j)}\right)\propto q\left(x_{i(j)}\right)$ which holds on the discrete support of the binary variable, i.e. $\left\{ -1,1\right\} $. In this notation, the dependency on $x_{i(j)}$ on the two distributions implies
that we are considering their marginal distributions. We now explicit write
the single-node marginals of variable $x_{i}$ w.r.t. $q^{\left(ij\right)}$ and $q$ (the same reasoning holds for $x_{j}$ by swapping the two indices):
\begin{equation}
q^{\left(ij\right)}\left(x_{i}\right)=\sum_{x_{j}=\pm 1}g^{\backslash ij}\left(x_{i},x_{j}\right)\psi_{ij}\left(x_{i},x_{j}\right)=\frac{1+x_{i}\left\langle x_{i}\right\rangle _{q^{\left(ij\right)}}}{2}\label{eq:DC_qtilted_a_singlemarg_i}
\end{equation}
where in the last equality the single-node marginal is written in
terms of its first moment $\left\langle x_{i}\right\rangle _{q^{\left(ij\right)}}$
(namely, the magnetization $m_i$), without loss of generality. In the same
spirit, the single node marginal of the full Gaussian distribution $q\left(\boldsymbol{x}\right)$
over node $i$ can be simply written as:

\begin{equation}
q\left(x_{i}\right)=\int d\boldsymbol{x}_{\backslash i}q\left(\boldsymbol{x}\right)=\frac{1}{\sqrt{2\pi\Sigma_{ii}}}\exp\left[-\frac{\left(x_{i}-\mu_{i}\right)^{2}}{2\Sigma_{ii}}\right]\label{eq:DC_q_singlemarg_i}
\end{equation}

Therefore, the matching of the density values of (\ref{eq:DC_qtilted_a_singlemarg_i})
and (\ref{eq:DC_q_singlemarg_i}) on $x_{i}=\left\{ -1,1\right\} $
can be rephrased by:

\begin{align}
q^{\left(ij\right)}\left(x_{i}\right)\propto q\left(x_{i}\right)\;\Longleftrightarrow\;\frac{q\left(x_{i}=+1\right)}{q\left(x_{i}=-1\right)} & =\frac{q^{\left(ij\right)}\left(x_{i}=+1\right)}{q^{\left(ij\right)}\left(x_{i}=-1\right)}\label{eq:DC_density_matching}
\end{align}

After some straightforward algebra, the following condition is obtained:

\begin{align}
\frac{\mu_{i}}{\Sigma_{ii}} & =\text{atanh}\left\langle x_{i}\right\rangle _{q^{\left(ij\right)}}\label{eq:DC_condition}
\end{align}

Eq. (\ref{eq:DC_condition}) is called \emph{DC condition} and it
is sufficient to ensure exact marginals computation on acyclic graphs: a rigorous proof can be found in the supplementary information of \citep{braunstein_loop_2019}. However, on loopy graphs it is necessary to complement the above set of equations with other conditions to fix all the parameters encoded by the Gaussian densities $\phi_{ij}$. The set of DC closures defined by \eqref{DCclosure_all} was chosen in \citep{braunstein_loop_2019} by imposing the matching of first moments between the marginal tilted and Gaussian distributions (Eq. \eqref{eq:DCclosure_mi}) and the matching of the Pearson correlation coefficient (Eq. \eqref{eq:DCclosure_sij}; in this case, DC condition (\ref{eq:DC_condition}) is used in Eq. \eqref{eq:DCclosure_sii} to fix the Gaussian variances $\Sigma_{ii}$.

The choice behind the matching of the Pearson correlation coefficient
is justified by imposing a unique transformation between the covariance
matrix of the (marginal) tilted distribution onto the Gaussian's one.
Indeed, defining the following map $\xi$
\begin{equation}
\xi\left(c_{ij},m_{i},m_{j}\right)=c_{ij}\sqrt{\frac{m_{i}}{\left(1-m_{i}^{2}\right)\text{atanh}m_{i}}}\sqrt{\frac{m_{j}}{\left(1-m_{j}^{2}\right)\text{atanh}m_{j}}}\label{eq:pearson_map_function}
\end{equation}

the closure equation for both diagonal and non-diagonal covariances
$\Sigma_{ij}$ can be rewritten as:

\begin{align}
\Sigma_{ij} & =\xi\left(\langle x_{i}x_{j}\rangle_{q^{\left(ij\right)}}-\langle x_{i}\rangle_{q^{\left(ij\right)}}\langle x_{j}\rangle_{q^{\left(ij\right)}},\langle x_{i}\rangle_{q^{\left(ij\right)}},\langle x_{j}\rangle_{q^{\left(ij\right)}}\right)\label{eq:pearson_map_covariances_all}
\end{align}

In particular, it is straightforward to show that, when $i=j$, the
map (\ref{eq:pearson_map_covariances_all}) reduces to Eq. (\ref{eq:DCclosure_sii})

On the other hand, the full moment matching closure inspired by EP implementations and given by Eqs. \eqref{EPclosure_all} is not exact on trees in general, as it does not satisfy Eq. \eqref{eq:DC_condition} (except when all the magnetizations are zero).

\section{Detailed computation of inferred parameters under DC scheme\label{sec:details_computation}}

We start by combining \eqref{Boltzmann_learning_all} with \eqref{tilted_moments_all}
and we solve w.r.t. $h_{i}^{\left(ij\right)},h_{j}^{\left(ij\right)},J_{ij}$:

\begin{align}
h_{i}^{\left(ij\right)*}= & \frac{1}{4}\log\frac{\left[\left(1+m_{i}\right)\left(1+m_{j}\right)+C_{ij}\right]\left[\left(1+m_{i}\right)\left(1-m_{j}\right)-C_{ij}\right]}{\left[\left(1-m_{i}\right)\left(1+m_{j}\right)-C_{ij}\right]\left[\left(1-m_{i}\right)\left(1-m_{j}\right)+C_{ij}\right]}-y_{i}^{c\left(ij\right)}=h_{i}^{(ij)IP}-y_{i}^{\left(ij\right)}\label{eq:tilted_inverse_hi}\\
h_{j}^{\left(ij\right)*}= & \frac{1}{4}\log\frac{\left[\left(1+m_{i}\right)\left(1+m_{j}\right)+C_{ij}\right]\left[\left(1-m_{i}\right)\left(1+m_{j}\right)-C_{ij}\right]}{\left[\left(1+m_{i}\right)\left(1-m_{j}\right)-C_{ij}\right]\left[\left(1-m_{i}\right)\left(1-m_{j}\right)+C_{ij}\right]}-y_{j}^{c\left(ij\right)}=h_{j}^{(ij)IP}-y_{j}^{\left(ij\right)}\label{eq:tilted_inverse_hj}\\
J_{ij}^{*}= & \frac{1}{4}\log\frac{\left[\left(1+m_{i}\right)\left(1+m_{j}\right)+C_{ij}\right]\left[\left(1-m_{i}\right)\left(1-m_{j}\right)+C_{ij}\right]}{\left[\left(1+m_{i}\right)\left(1-m_{j}\right)-C_{ij}\right]\left[\left(1-m_{i}\right)\left(1+m_{j}\right)-C_{ij}\right]}+S^{c\left(ij\right)}=J_{ij}^{IP}+S^{\left(ij\right)}\label{eq:tilted_inverse_Jij}
\end{align}

The terms $h_{i}^{(ij)IP}$, $h_{j}^{(ij)IP}$,$J_{ij}^{IP}$ are
the inferred parameters in the context of the Independent Pair approximation.
In particular, $h_{i}^{\left(ij\right)}$represents the contribution
to the inferred field on spin $i$ when considered in pair with $j$.

The expression of couplings is derived directly from (\ref{eq:tilted_inverse_Jij})
by inserting the expression for of the cavity coupling (\ref{eq:cav_coupling_Sij}):

\[
J_{ij}^{*}=J_{ij}^{IP}-\left(\boldsymbol{\Sigma}^{-1}\right)_{ij}-\frac{\Sigma_{ij}}{\Sigma_{ii}\Sigma_{jj}-\Sigma_{ij}^{2}}\qquad\forall i\neq j
\]

The overall external field acting on spin $i$ will be computed as
$h_{i}^{*}=\sum_{j\neq i}h_{i}^{\left(ij\right)}$. By using (\ref{eq:tilted_inverse_hi})
and the expression of cavity fields (\ref{eq:cav_field_ij}) we get:

\begin{align}
h_{i}^{*} & =\sum_{j\neq i}h_{i}^{\left(ij\right)}=\sum_{j\neq i}\left(h_{i}^{(ij)IP}-\frac{\Sigma_{jj}m_{i}-\Sigma_{ij}m_{j}}{\Sigma_{ii}\Sigma_{jj}-\Sigma_{ij}^{2}}\right)+\sum_{j\neq i}\gamma_{i}^{\left(ij\right)}\\
 & =\sum_{j\neq i}h_{i}^{(ij)IP}-\sum_{j\neq i}\frac{\Sigma_{jj}m_{i}-\Sigma_{ij}m_{j}}{\Sigma_{ii}\Sigma_{jj}-\Sigma_{ij}^{2}}+\left(\boldsymbol{\Sigma}^{-1}\boldsymbol{m}\right)_{i}\\
 & =h_{i}^{IP}+\left(N-2\right)\text{atanh}m_{i}-\sum_{j\neq i}\frac{\Sigma_{jj}m_{i}-\Sigma_{ij}m_{j}}{\Sigma_{ii}\Sigma_{jj}-\Sigma_{ij}^{2}}+\left(\boldsymbol{\Sigma}^{-1}\boldsymbol{m}\right)_{i}
\end{align}

where in the second line we used $\left(\boldsymbol{\Sigma}^{-1}\boldsymbol{m}\right)_{i}=\sum_{j\neq i}\gamma_{i}^{\left(ij\right)}$
as in (\ref{eq:gauss_moments_wrt_params}).

\section{Reconstruction with Linear-Response methods}

We report here the expression of inferred fields and couplings by means of Mean-Field, TAP and Susceptibility Propagation (SP) approximations. All of them allow to estimate of couplings by means of Linear Response relations, computed w.r.t. to MF, TAP, and BP fixed points, respectively. We refer to \citep{SP_Ricci_Tersenghi,nguyen_inverse_2017} for a detailed
derivation of all them.
\subsubsection{Naif Mean Field}

\begin{align}
J_{ij}^{MF} & =-\left(\boldsymbol{C}^{-1}\right)_{ij}\label{eq:Jij_MF}\\
h_{i}^{MF} & =\text{atanh}m_{i}-\sum_{j}J_{ij}^{MF}m_{j}\label{eq:hi_MF}
\end{align}

\subsubsection{TAP}
\begin{align}
J_{ij}^{TAP} & =\frac{\sqrt{1-8m_{i}m_{j}\left(\boldsymbol{C}^{-1}\right)_{ij}}-1}{4m_{i}m_{j}}\label{eq:Jij_TAP}\\
h_{i}^{TAP} & =\text{atanh}m_{i}-\sum_{j}J_{ij}^{TAP}m_{j}+m_{i}\sum_{j\neq i}\left(J_{ij}^{TAP}\right)^{2}\left(1-m_{j}^{2}\right)\label{eq:hi_TAP}
\end{align}

\subsubsection{Susceptibility Propagation (SP)}
\begin{multline}
J_{ij}^{SP}=-\text{atanh}\left[\frac{1}{2\left(\boldsymbol{C}^{-1}\right)_{ij}}\sqrt{1+4\left(1-m_{i}^{2}\right)\left(1-m_{j}^{2}\right)\left(\boldsymbol{C}^{-1}\right)_{ij}^{2}}-m_{i}m_{j}-\right.\\
\left.\frac{1}{2\left(\boldsymbol{C}^{-1}\right)_{ij}}\sqrt{\left(\sqrt{1+4\left(1-m_{i}^{2}\right)\left(1-m_{j}^{2}\right)\left(\boldsymbol{C}^{-1}\right)_{ij}^{2}}-2m_{i}m_{j}\left(\boldsymbol{C}^{-1}\right)_{ij}\right)^{2}-4\left(\boldsymbol{C}^{-1}\right)_{ij}^{2}}\;\right]\label{eq:Jij_SP}
\end{multline}

\begin{equation}
h_{i}^{SP}=\text{atanh}m_{i}-\sum_{j\neq i}\text{atanh}\left[t_{ij}f\left(m_{j},m_{i},t_{ij}\right)\right]
\end{equation}

where $t_{ij}=\text{tanh}J_{ij}$ and
\begin{equation}
f\left(m_{1},m_{2},t\right)=\frac{\left(1-t^{2}\right)-\sqrt{\left(1-t^{2}\right)^{2}-4t\left(m_{1}-tm_{2}\right)\left(m_{2}-tm_{1}\right)}}{2\left(m_{2}-tm_{1}\right)}
\end{equation}
\end{document}